\documentclass[11pt]{article}

\pdfoutput=1
\usepackage{amsmath}
\usepackage{amsthm}
\usepackage{amsfonts}          
\usepackage{amssymb}           
\usepackage{color}
\usepackage{mathtools}
\usepackage{cite}
\usepackage{multirow,setspace,microtype}
\usepackage{bm}
\usepackage{booktabs}
\usepackage{tikz}
\usepackage{slashed}

\definecolor{darkblue}{rgb}{0.0,0.0,0.3} 		

\usepackage[colorlinks]{hyperref}
\hypersetup{pdftitle={EffectiveTheory},
	pdfauthor={ Sebastian Dumitru and Burt Ovrut},
	linkcolor=darkblue, urlcolor=darkblue, citecolor=darkblue, filecolor=black, linktoc=page}

\PassOptionsToPackage{linecolor=blue,backgroundcolor=blue!25,bordercolor=blue,textsize=scriptsize}{todonotes}
\usepackage{todonotes}

\numberwithin{equation}{section}

\usepackage{makecell}

\usepackage[bf]{caption}
\let\originalleft\left
\let\originalright\right
\renewcommand{\left}{\mathopen{}\mathclose\bgroup\originalleft}
\renewcommand{\right}{\aftergroup\egroup\originalright}

\makeatletter
\g@addto@macro\bfseries{\boldmath}
\makeatother

\newlength{\xtrawidth}
\newlength{\xtraheight}
\usepackage{subcaption}
\usepackage{float}
\usepackage{caption}
\allowdisplaybreaks

\begin{document}

\begin{titlepage}
\title{\LARGE {Heterotic $M$-Theory Hidden Sectors with an Anomalous $U(1)$ Gauge Symmetry}\\[.3cm]}
                       
\author{{
   Sebastian Dumitru
   and Burt A.\,Ovrut}\\[0.8cm]
   {${}$\it Department of Physics, University of Pennsylvania} \\[.1cm]
   {\it Philadelphia, PA 19104, USA}
 }      

\date{}

\maketitle

\begin{abstract}
\noindent
The hidden sector of heterotic $M$-theory vacua whose gauge bundle contains an anomalous $U(1)$ factor is discussed in detail.  The mathematical formalism for computing the inhomogeneous transformation of the dilaton and K\"ahler moduli axions under an anomalous $U(1)$ transformation is presented.  Along with matter multiplets, which transform homogeneously under $U(1)$, the relevant part of the $U(1)$ invariant low energy hidden sector Lagrangian is presented and analyzed. A detailed mathematical formalism is given for rotating these field to a new basis of chiral superfields with normalized kinetic energy and a diagonal mass matrix. It is shown that the scalar and fermionic components of one such chiral superfield give rise to a massive $U(1)$ vector superfield, whose mass is composed of both anomalous and Higgs contributions associated with the inhomogeneous and homogeneous transformations respectively. Two explicit examples are presented, the first with vanishing and the second with non-zero Fayet-Iliopoulos term. The canonically normalized hidden sector Lagrangian given here is restricted to gauge interactions only. A study of higher order interactions of the moduli and matter multiplets, as well as the coupling to supergravity, will be presented elsewhere.

\noindent

\let\thefootnote\relax\footnotetext{\noindent sdumitru@sas.upenn.edu, ovrut@elcapitan.hep.upenn.edu}

\end{abstract}

\thispagestyle{empty}
\end{titlepage}

\tableofcontents

\section{Introduction}

Heterotic $M$-theory, first introduced in \cite{Lukas:1997fg} and discussed in detail in \cite{Lukas:1998yy, Lukas:1998tt}, is eleven-dimensional Horava-Witten theory
 \cite{Horava:1995qa,Horava:1996ma} dimensionally reduced to five-dimensions by compactifying on a Calabi-Yau (CY) complex
threefold. The five-dimensional heterotic $M$-theory consists of two four-dimensional orbifold planes separated by a finite fifth-dimension; specifically, a real one-dimensional manifold of the form $S^{1}/{\mathbb{Z}}_{2}$. The two orbifold planes, each
with an $E_{8}$ gauge group, are called the observable and hidden sectors respectively~\cite{Lukas:1997fg, Lukas:1998yy, Lukas:1998tt, Donagi:1998xe,Ovrut:2000bi}. By choosing a suitable CY threefold, as well as an
appropriate holomorphic vector bundle~\cite{Donagi:1999gc} on the CY compactification at the observable sector, one can find realistic low energy $N=1$ supersymmetric particle physics models. A number of such realistic observable sector theories
have been constructed. See, for example,~\cite{Braun:2005nv,Braun:2005bw,Braun:2005ux,Bouchard:2005ag,Anderson:2009mh,Braun:2011ni,Anderson:2011ns,Anderson:2012yf,Anderson:2013xka,Nibbelink:2015ixa,Nibbelink:2015vha,Braun:2006ae,Blaszczyk:2010db,Andreas:1999ty,Curio:2004pf}.
Similarly, by choosing a holomorphic gauge bundle at the hidden wall, one can obtain a variety of low energy hidden sector particle physics theories. It is essential, however, that having specified the observable sector bundle, the hidden sector bundle must satisfy a number of specific, and non-trivial, constraints. These constraints have been specified in detail in \cite{Braun:2013wr,Ovrut:2018qog,Ashmore:2020ocb}, and greatly reduce the allowed low energy theories of the hidden sector. Appropriate hidden sector vector bundles, and their low energy particle physics theories, have been studied far less frequently and with less detail. See, for example, \cite{Anderson:2009nt,Anderson:2011cza,Nilles:1998sx,Lukas:1997rb,Blumenhagen:2006ux,Weigand:2006yj,Ashmore:2020wwv}. 
Several of these admissible hidden sectors have gauge bundles which contain an ``anomalous'' $U(1)$ factor \cite{Dine:1987xk,Dine:1987gj,Anastasopoulos:2006cz}. 
For example, this has been accomplished within the context of the so-called $B-L$ MSSM theory ~\cite{Ambroso:2009jd,Marshall:2014kea,Marshall:2014cwa,Ovrut:2012wg,Ovrut:2014rba,Barger:2008wn,FileviezPerez:2009gr}, using only a single line bundle in the hidden sector \cite{Ashmore:2020ocb,Ashmore:2020wwv}. Such low energy hidden sector theories can, in principle, contain important new physics; for example, having particles in their spectrum that can act as candidates for ``dark matter'' in cosmological theories. It follows that a detailed study of the hidden sector spectrum and interactions in realistic heterotic $M$-theory vacua is potentially important. That is the motivation for the present paper. For specificity, we will work within the context of hidden sector gauge bundles that contain a single anomalous $U(1)$ factor in their structure group.

Although various aspects of anomalous $U(1)$ properties have been studied, they are usually within the context of the observable sector \cite{Ibanez:2001nd,Aldazabal:2000dg}, or relatively specific contexts that are not directly hidden sectors 
in a realistic heterotic $M$-theory vacuum \cite{Blumenhagen:2005ga,Blumenhagen:2006ux,Weigand:2006yj,Lukas:1999nh,Anderson:2009nt,Anderson:2010mh,Binetruy:1996uv}\footnote{In\cite{Lalak:1998jg} the authors study yet a different scenario, in which each sector contains an anomalous $U(1)$.}. 
 Here, we will construct the general mathematical structure of realistic hidden sectors with an anomalous $U(1)$ in the structure group. Specifically, the general formalism for both the {\it inhomogeneous} anomalous transformations of the axions associated with both the dilaton and the K\"ahler moduli of an arbitrary CY threefold will be derived and added to the usual homogeneous transformations of the low energy matter fields of the hidden sector compactification. The relevant portion of the $U(1)$ invariant low energy effective Lagrangian will be specified, as will the associated K\"ahler potentials, K\"ahler metrics  and Killing vectors of the moduli and matter fields. It will then be shown that, using a specific redefinition of fields, a Lagrangian with canonically normalized kinetic energy terms as well as a diagonal mass matrix can be produced. Using this formalism, we can compute the general anomalous and Higgs mass of the $U(1)$ gauge multiplet and, having done so, specify the effective relevant Lagrangian for the remaining low mass chiral superfields. 

Specifically, we do the following. In Section 2, we give a brief summary of a) the analysis of the Green-Schwarz mechanism~\cite{Green:1984sg} that leads to inhomogeneous transformations of the axions associated with the dilaton and K\"ahler moduli scalar fields and b) present the effective Lagrangian for the dilaton, moduli and matter chiral superfields--as well as the gauge connection and associated gaugino--invariant under the spontaneously broken anomalous $U(1)$ super-gauge transformation. The result of rotating these fields to a  specific basis with canonical kinetic energy and diagonal mass terms is presented and analyzed. In Section 3, we take a specific vacuum in which the Fayet-Iliopoulos term vanishes and, hence, the low energy matter fields do not have vacuum expectation values. The physical ideas presented in Section 2 now greatly simplify and we discuss the results in detail. 
The formalism for another specific vacuum, but now with a non-vanishing Fayet-Iliopoulos term, is presented and analyzed in Section 4. In this case, the matter scalars have non-vanishing vacuum expectation values and, hence, the results are more complicated, Following an initial discussion, we simply present the relevant final results.
In Appendex A, we give a detailed analysis of the Green-Schwarz mechanism in both ten-dimensions, its dimensional reduction to four-dimensions and a derivation of the associated inhomogenous transformation of the dilaton and K\"ahler moduli axions. Finally, in Appendix B we present a detailed mathematical analysis of the the relevant portion of the effective Lagrangian involving chiral superfields, as well as the gauge connection and gaugino,  that is invariant under an anomalous $U(1)$ gauge transformation. The superfield rotation to a canonically normalized Lagrangian with a massive $U(1)$ vector superfield as well as massless chiral moduli/matter superfields is presented in detail. 

The formalism presented in this paper will be used, in conjunction with heterotic $M$-theory cosmological theories, such as \cite{Deen:2016zfr,Cai:2018ljy,Ibanez:2014swa}, to explore potentially new types of dark matter--composed of hidden sector moduli~\cite{Chowdhury:2018tzw,Dutra:2019nhh} and matter fields, both scalars and fermions. This work will be presented elsewhere.

\section{Anomalous $U(1)$ Hidden Sectors}

We consider ten-dimensional Horava-Witten theory \cite{Horava:1995qa, Horava:1996ma} compactified to five-dimensional heterotic M-theory vacua on a Calabi-Yau threefold $X$. The compactified theory consists of two four-dimensional orbifold planes, the  observable and hidden  sectors respectively, separated by a finite one-dimensional interval of the form $S^{1}/ \mathbb{Z}_{2}$. In this paper, we will focus on the hidden sector only.

\subsection{Hidden Sector Gauge Bundle}

The hidden sector gauge bundle has a structure group of the form
\begin{equation}
G^{(2)}=\mathcal{G}^{(2)}\times U(1)\ ,
\end{equation}
where $U(1)$ is the structure group of a line bundle $L$ appropriately embedded into the hidden sector $E_8$ gauge group. If $h^{1,1}$ is the number of harmonic $(1,1)$ forms on $X$, then $L$ is defined as 
\begin{equation}
L=\mathcal{O}_X(l^1,\dots, l^{h^{1,1})} \ ,
\end{equation}
where $l^i$, $i=1,\dots, h^{1,1}$ are integers.  The $\mathcal{G}^{(2)}$ is the structure group of a possible additional bundle factor. This $\mathcal{G}^{(2)}$ factor will not play a role in our analysis and, henceforth, we will ignore it. The gauge group of the hidden sector low energy effective theory is then the commutant of $U(1)$ in $E_8$ of the form
\begin{equation}
H^{(2)}=\mathcal{H}^{(2)}\times U(1)\ .
\end{equation}
Here $\mathcal{H}^{(2)}$ depends on the explicit embedding of $U(1)$ into $E_8$. Of course, $U(1)\subset H^{(2)}$ since $U(1)$ commutes with itself. We will denote the $U(1)$ gauge connection and its Weyl spinor gaugino superpartner by $A_{2\mu}$ and $\lambda_{2}$ respectively. The spectrum of chiral superfields in the low energy effective theory is the following.

\subsection{Moduli Chiral Superfields}

We first discuss the  superfields associated with the moduli of the Calabi-Yau threefold $X$, the modulus of the fifth-dimension $S^1/\mathbb{Z}_2$ interval, and the modulus associated with the position of an internal five-brane. There are two types of moduli for the threefold $X$; namely, the K\"ahler and complex structure moduli respectively. However, as shown in~\cite{Anderson:2010mh, Anderson:2011cza}, the complex structure moduli are always stabilized at the compactification scale and, therefore, are absent from the 4D low energy effective theory. Henceforth, we will ignore them.  The relevant K\"ahler moduli $a^{i}, i=1,\dots,h^{1,1}$, the $S^1/\mathbb{Z}_2$ interval modulus $\hat{R}$ and internal five-brane modulus $z=\tfrac{1}{2}+\lambda$ combine to form the real part of the scalar components of three types of chiral superfields; namely the dilaton $\tilde S$,the K\"ahler superfields $\tilde T^i$ and the five brane moduli $\tilde Z$. Following~\cite{Brandle:2003uya,Anderson:2009nt} (see also~\cite{Weigand:2006yj} for the weakly coupled string result and \cite{Moore:2000fs} for a detailed derivation), the expressions for the complex scalar components of these superfields, in the presence of five-branes, are
\begin{equation}
\begin{split}
\label{eq:def_scalar_intro}
& S=V+\pi\epsilon_SW_it^iz^2
+i\left[\sigma+2\pi \epsilon_S W_i\chi^iz^2\right] \ ,\\
&T^i=t^i+2i\chi^i\ ,\quad i=1,\dots,h^{1,1}\ ,\\
&Z=W_it^iz+2iW_i(-\eta^i\nu+\chi^iz)\ ,
\end{split}
\end{equation}
where 
\begin{equation}
V=\frac{1}{6}d_{ijk}a^{i}a^{j}a^{k} , \quad t^i=\frac{\hat R a^i}{V^{1/3}}\ ,
\end{equation}
$W_i$ specifies properties of the internal five-brane and $d_{ijk}$ are the intersection numbers for $X$. Note that the complex components of $S$, $T^{i}$ and $Z$ contain axionic scalars $\sigma$, $\chi^{i}$ and $\eta^i$, for $i=1,\dots,h^{1,1}$, respectively, in addition to the $a^{i}$, $\hat{R}$ and $z=\tfrac{1}{2}+\lambda$ moduli presented above. All of these moduli, axions and parameters were defined and discussed in detail in \cite{Ashmore:2020ocb}. 
Finally, as presented in~\cite{Brandle:2003uya,Anderson:2009nt}, the K\"ahler potentials for the scalar components of these moduli are
\begin{equation}
\begin{split}
\label{eq:K_SandK_T}
&K_S=-\kappa_4^{-2}\ln\left(S+\bar S-\frac{\pi}{2}\epsilon_S \frac{(Z+\bar Z)^2}{W_i(T^i+\bar T^i)}\right)\ , \\
&K_T=-\kappa_4^{-2}\ln\left(\frac{1}{48}d_{ijk}(T^i+\bar T^i)(T^j+\bar T^j)(T^k+\bar T^k)\right)\ ,\\
\end{split}
\end{equation}

It is important to note that the moduli superfields are uncharged under 
the low energy $U(1)$ gauge group and, hence, do {\it not} transform homogeneously under U(1).  However, as we will show, they
do transform {\it inhomogenously} under an anomalous $U(1)$ gauge transformation.

\subsection{Matter Chiral Superfields}

In addition to the dilaton and K\"ahler moduli, the hidden sector low energy effective theory contains matter chiral superfields whose fermions are associated with the zero-modes of the 6D Dirac operator. These can be determined, for example, using the Euler characteristic of powers of the associated bundle, as discussed in~\cite{Green:1987mn}. Unlike the moduli superfields, these matter superfields generically transform {\it homogeneously} under both the $\mathcal{H}^{(2)}$ and 
$U(1)$ subfactors of the low energy gauge group. In this paper, for simplicity, we restrict our discussion to the fields which transform non-trivially under $U(1)$, but are ${\cal{H}}^{(2)}$ singlets. We specify such matter superfields as $C^L$, $L=1,\dots,\mathcal{N}$, and denote their $U(1)$ charges by $Q^L$. That is, under an infinitesimal $U(1)$ gauge transformation
\begin{equation}
\label{eq:def_matter_transform_intro}
\delta_\theta C^L=-iQ^LC^L\theta\equiv k_C^L \theta\ , \quad L=1, \dots, \mathcal{N}\ ,
\end{equation}
where $\theta=\theta(x)$ is an infinitesimal parameter. Finally, as presented in~\cite{Brandle:2003uya,Lukas:1998tt}, the K\"ahler potential associated with the matter superfields is given by
\begin{equation}
\label{eq:K_matter_int}
K_{\text{matter}}=e^{ \kappa^2_4K_T/3}\mathcal{G}_{L\bar M}C^L\bar C^{\bar M}\ ,
\end{equation}
where $\mathcal{G}_{L\bar M}$ is an unspecified, generically moduli dependent, Hermitian matrix on the $H^1$ cohomologies associated with the $C^L$ matter fields in the hidden sector. In this paper, we will, for simplicity, assume that $\mathcal{G}_{L\bar M}$ is a constant matrix.

\subsection{Anomalous $U(1)$}

Importantly, we henceforth assume that the line bundle $L$ and its embedding into $E_8$ are chosen so that the charges $Q^L$ lead to an anomalous gauge 3-point function. As is well-known, in string theory this anomaly can be removed by the Green-Schwarz mechanism~\cite{Green:1984sg}. The associated $U(1)$ structure group of $L$ is then referred to as an ``anomalous'' $U(1)$ and has a number of important properties. The most relevant for this paper is the fact that the Green-Schwarz mechanism induces
an {\it inhomogenous} transformation in both the dilaton, the K\"ahler moduli and the five-brane modulus discussed above, even though they are uncharged under $U(1)$. The Green-Schwarz mechanism for an anomalous $U(1)$ and the resulting inhomogenous transformations for the dilaton, K\"ahler moduli and five-brane mudulus are presented in detail in Appendix A. Here we simply state the results.
We found that under an infinitesimal anomalous $U(1)$ gauge transformation, the scalar parts of these moduli transform as

\begin{equation}
\begin{split}
\label{eq:def_scalar_transform_intro}
& \delta_\theta S=-2i\pi a\epsilon_S^2\epsilon_R^2\left(\tfrac{1}{2} \beta^{(2)}_i l^i + W_il^iz^2\right)\theta\ \equiv k_S\theta,\\
&\delta_\theta T^i=-2i a\epsilon_S\epsilon_R^2l^i\theta \equiv k_T^i\theta\ ,\quad i=1,\dots,h^{1,1}\ ,\\
&\delta_\theta Z=-2ia\epsilon_S\epsilon_R^2W_il^iz\theta=k_Z\theta\ .\\
\end{split}
\end{equation}
The $U(1)$ homogenous gauge transformation of the matter scalars $C^L$, $L=1,\dots, \mathcal{N}$ was presented in \eqref{eq:def_matter_transform_intro} and remains unchanged for an anomalous $U(1)$. That is, under an anomalous $U(1)$ gauge transformation 
\begin{equation}
\label{eq:def_matter_transform_introB}
\delta_\theta C^L=-iQ^LC^L\theta\equiv k_C^L \theta\ , \quad L=1, \dots, \mathcal{N}\ .
\end{equation}
The explicit vectors $k_S$, $k_T^i$, $k_{Z}$ and $k_C^L$, defined by these transformations, that is
\begin{equation}
\begin{split}
& k_S=-2i\pi a\epsilon_S^2\epsilon_R^2\left(\tfrac{1}{2} \beta^{(2)}_i l^i + W_il^iz^2\right) \ ,\\
& k_T^i =-2i a\epsilon_S\epsilon_R^2l^i\ ,\\
&k_Z=-2ia\epsilon_S\epsilon_R^2W_il^iz\ ,\\
& k_C^L=-iQ^LC^L
\end{split}
\label{rain2}
\end{equation}
can be shown to be the Killing vectors on the space of all moduli and the matter chiral superfields.

\subsection{Hidden Sector Lagrangian}

Having presented all the moduli and the relevant matter chiral superfields in the hidden sector, along with their transformation
properties under the anomalous $U(1)$ gauge group, one can now present the complete Lagrangian of the low energy effective field theory. We begin by explicitly writing all these chiral superfields in terms of their component fields as 
\begin{equation}
\begin{split}
&\tilde S=(S,\psi_S,F_S)\ ,\\
&\tilde T^i=(T^i,\psi_T^i,F_T^i),\quad i=1,\dots, h^{1,1}\ ,\\
&\tilde Z=(Z,\psi_Z,F_Z)\ ,\\
&\tilde C^L=(C^L,\psi^L,F^L), \quad L=1,\dots, \mathcal{N}\ .
\end{split}
\end{equation}
The transformations of the scalar components of these superfields under the anomalous $U(1)$ were presented in the previous section. Considering these transformations, one defines the following covariant derivatives for the scalar components 
\begin{equation}
\begin{split}
&{D}_\mu S=\partial_\mu S-A_\mu  k_S\ ,\\
&D_\mu T^i=\partial_\mu T^i-A_\mu k_T^i\ ,\quad i=1,\dots,h^{1,1}\ ,\\
&D_\mu Z=\partial_\mu Z-A_\mu k_Z\ ,\\
&D_\mu C^L=\partial_\mu C^L-A_\mu k_C^L\ ,\quad L=1,\dots, \mathcal{N}\ .\\
\end{split}
\end{equation}

Finally, we recall that the complete K\"ahler potential for $S$, $T^i$, $Z$ and $C^L$ fields is given by
\begin{equation}
\label{blue1}
K=K_S+K_T+K_{\text{matter}}\ ,
\end{equation}
where $K_S$, $K_T$ and $K_{\text{matter}}$ were given in \eqref{eq:K_SandK_T} and \eqref{eq:K_matter_int}. 
Given all this information, one can now write the most general 4D Lagrangian for the superfields $\tilde{S}, \tilde{T}^i$, $\tilde{Z}$ and $\tilde{C}^L$ in the hidden sector that is invariant under anomalous $U(1)$ gauge transformations. The relevant terms in the four-dimensional effective action are
{\small
\begin{equation}
\label{rain1}
\begin{split}
\mathcal{L}\supset&- g_{S\bar S}D_\mu S D^\mu \bar S-g_{T^i\bar T^j}D_\mu T^i D^\mu \bar T^{\bar j}-g_{Z\bar Z}D_\mu Z D^\mu \bar Z- g_{C^L\bar C^{\bar M}}D_\mu C^L D^\mu \bar C^{\bar M}\\
&(- g_{S\bar T^i}D_\mu S D^\mu \bar T^i- g_{S\bar Z}D_\mu S D^\mu \bar Z- g_{T^i\bar C^L}D_\mu T^i D^\mu \bar C^L
 - g_{Z\bar T^i}D_\mu Z D^\mu \bar T^i+hc)\\
&+\left[ -ig_{S\bar S}\psi_S \slashed{\mathcal{D}} \psi_S^{\dag}
 -ig_{T^i\bar T^j}\psi_T^i \slashed{\mathcal{D}} \psi_T^{\bar j\dag}  -ig_{Z\bar Z}\psi_Z \slashed{\mathcal{D}} \psi_Z^{\dag}-ig_{C^L\bar C^{\bar M}}\psi^L \slashed{\mathcal{D}} \psi^{\bar M\dag}+\cdots\right]\\
&+\sqrt{2}\Big[g_{S\bar S}k_S\lambda_{2}^\dag \psi_S+g_{S\bar S}\bar k_S\lambda_{2} \psi_S+g_{T^i\bar T^j}k_T^i\lambda_{2}^\dag \psi_T^{\bar j\dag}+g_{T^i\bar T^j}\bar k_T^{\bar j}\lambda_{2} \psi_T^i\\
&\qquad\qquad+g_{Z\bar Z}k_Z\lambda_{2}^\dag \psi_Z+g_{Z\bar Z}\bar k_Z\lambda_{2} \psi_Z+g_{C^L\bar C^{\bar M}}k_C^L\lambda_{2}^\dag \psi^{\bar M\dag}+g_{C^L\bar C^{\bar M}}\bar k_C^{\bar M}\lambda_{2} \psi_C^L+\dots\Big]\\
 &-\frac{i}{g_2^2} \lambda_{2} \slashed{\partial} \lambda_{2}^\dag
-\frac{1}{4g_2^2}F^{\mu \nu}_2 F_{2\mu \nu}-\frac{g_2^2}{2}\mathcal{P}^2-g_{C^L\bar C^{\bar M}}\frac{\partial \mathcal W}{\partial C^L} \frac{\partial \bar{\mathcal{W}}}{\partial \bar C^{\bar M}}\ .\\
\end{split}
\end{equation}
}
In the above expression, we show all the non-zero scalar kinetic terms, including ``cross-terms'' of the type $g_{S\bar T^i}D_\mu S D^\mu \bar T^i$. The dots represent cross-terms in the fermion kinetic functions, as well as in their couplings to the gaugino field, which we omit for brevity. They are completely analogous to the scalar cross-terms, however.
In Appendix C we give the expressions  for the K\"ahler metrics--$g_{S\bar S}$, $g_{T^i\bar T^j}$, $g_{Z\bar Z}$, $g_{C^L\bar C^{\bar M} }$, $g_{S\bar T^i}$, $g_{S\bar Z}$, $g_{T^i\bar C^{\bar L}}$, $g_{Z\bar T^i}$-- obtained after differentiating the K\"ahler potential with respect to the fields $S$, $T^i$, $Z$ and $C^L$. 

$F_2^{\mu \nu}=\partial^\mu A_2^\nu-\partial^\nu A_2^\mu$ is the field strength associated with the anomalous $U(1)$ on the hidden sector and $g_2$ is the $U(1)$ gauge coupling on the hidden sector, given by\cite{Ashmore:2020ocb}
\begin{equation}
g_2^2=\frac{\pi \hat \alpha_{\text{GUT}}}{a\text{Re}f_2}\ .
\end{equation}
%
It follows from expression \eqref{rain1} that the scalar potential energy is given by 
\begin{equation}
V=V_D+V_F\ ,
\end{equation}
where the $D$ and $F$ terms are
\begin{equation}
V_D=\frac{1}{2}g_2^{2}\mathcal{P}^2\ ,\quad V_F=e^{\kappa_4^2K_T/3}\mathcal{G}_{L\bar M}\frac{\partial \mathcal W}{\partial C^L} \frac{\partial \bar{\mathcal{W}}}{\partial \bar C^{\bar M}}
\end{equation}
respectively. First consider $V_D$. The moment map $\mathcal{P}$ is defined by
\begin{equation}
\mathcal{P}=ik_S\frac{\partial K}{\partial S}+ik_T^i\frac{\partial K}{\partial T^i}+ik_Z\frac{\partial K}{\partial Z}
+ik_C^L\frac{\partial K}{\partial C^L}\ .
\label{lab1}
\end{equation}
Using \eqref{blue1}, we have
\begin{equation}
\begin{split}
\frac{\partial K}{\partial S}=&\frac{\partial K_S}{\partial S}=-\frac{1}{2\kappa^2_4V}\ ,\\
\frac{\partial K}{\partial T^i}=& \frac{\partial K_T}{\partial T^i}+\frac{\partial K_S}{\partial T^i} +\frac{\partial K_{\text{matter}}}{\partial T^i}\\
=&-\frac{d_{ijk}a^ja^k}{4\kappa_4^2\hat RV^{2/3}}-\frac{1}{4\kappa_4^2V}\pi\epsilon_Sz^2W^i-\frac{d_{ijk}a^ja^k}{4\hat RV^{2/3}}\frac{e^{\kappa_4^2K_T/3}}{3}\mathcal{G}_{L\bar M}C^L\bar C^{\bar M}\ ,\\
\frac{\partial K}{\partial Z}=&\frac{\partial K_S}{\partial Z}=\frac{\pi\epsilon_Sz}{2\kappa^2_4V}\ ,\\
\frac{\partial K}{\partial C^L}=&\frac{\partial K_{\text{matter}}}{\partial C_L}=e^{\kappa_4^2K_T/3}\mathcal{G}_{L\bar M}\bar C^{\bar M}\ .
\end{split}
\end{equation} \ .
Therefore, using the Killing vector expressions from \eqref{rain2}, we find that
\begin{equation}
\begin{split}
\label{rain3}
\mathcal{P}&={-\frac{a\epsilon_S\epsilon_R^2}{2\kappa_4^2\hat RV^{2/3}}\left( \mu(L)+\frac{\pi \epsilon_S\hat R}{V^{1/3}}\left(\beta_i^{(2)}+z^2W_i\right)l^i    \right)}\\
&\qquad\qquad+\mathcal{G}_{L\bar M} e^{\kappa_4^2K_T/3}{\left(   1+   \frac{a\epsilon_S\epsilon_R^2}{12Q^L\hat RV^{2/3}} {\mu(L)}\right)}  Q^L C^L\bar C^{\bar M}\\
&={-\frac{a\epsilon_S\epsilon_R^2}{2\kappa_4^2\hat RV^{2/3}}\left( \mu(L)+\frac{\pi \epsilon_S\hat R}{V^{1/3}}\left(\beta_i^{(2)}+z^2W_i\right)l^i    \right)} + G_{L\bar M} C^L\bar C^{\bar M}
\end{split}
\end{equation}
where 
\begin{equation}
\mu(L)=d_{ijk}l^ia^ja^k 
\label{rain4A}
\end{equation}
is the tree level expression for the slope of the line bundle $L$ and we have defined the metric
\begin{equation}
\label{def_GLM}
G_{L\bar M}=\mathcal{G}_{L\bar M} e^{\kappa_4^2K_T/3}{\left(   Q^L+   \frac{a\epsilon_S\epsilon_R^2}{12\hat RV^{2/3}} {\mu(L)}\right)} \ .
\end{equation}

Next, consider $V_F$. The perturbative superpotential $\mathcal{W}$ can be shown to be independent of the dilaton and the K\"ahler moduli~\cite{Lukas:1998yy,Lukas:1997fg,Polchinski:1998rr}. As a result, only terms proportional to $\frac{\partial \mathcal{W}}{\partial C^L}$ are potentially non-vanishing. However, one cannot generically form a gauge invariant superpotential out of the matter fields $C^L$ which produce a $U(1)$
anomaly, as in our case. Therefore, our hidden sector does not contain a perturbative contribution to the superpotential and, hence,
\begin{equation}
V_F=e^{\kappa_4^2K_T/3}\mathcal{G}_{L\bar M}\frac{\partial \mathcal W}{\partial C^L} \frac{\partial \bar{\mathcal{W}}}{\partial \bar C^{\bar M}}=0\ .
\end{equation}
It follows that the scalar potential of the effective theory is simply
\begin{equation}
V=V_D=\frac{1}{2}g_2^{2}\mathcal{P}^2\ .
\label{black2}
\end{equation}

\subsection{Supersymmetric Vacua}

In order for the vacuum to be $N=1$ supersymmetric, it is necessary for the vacuum expectation values of the fields $S$, $T^i$, $Z$ and $C^L$ to satisfy the D-flatness condition
\begin{equation}
V=V_D=0\quad \Rightarrow \quad \langle \mathcal{P}  \rangle =0
\end{equation} 
or equivalently, from \eqref{lab1}, that
\begin{equation}
i\langle k_S\frac{\partial K}{\partial S}\rangle+i\langle k_T^i\frac{\partial K}{\partial T^i}\rangle+i\langle k_Z\frac{\partial K}{\partial Z}\rangle+i \langle k_C^L\frac{\partial K}{\partial C^L}\rangle =0 \ .
\label{rain4}
\end{equation}
It is important to note from \eqref{rain2} that $k_{S}$, $k_Z$ and $k_{C}^{L}$ are functions of the modulus $z=\tfrac{1}{2}+\lambda$ and the matter fields $C^{L}$ respectively and, hence, these parameters must be evaluated at their vacuum expectation values as well.  

At this point, it is useful to define the so-called Fayet-Iliopoulos (FI) term within our present context. The FI term is defined to be
\begin{align}
FI&=i\langle k_S\frac{\partial K}{\partial S}\rangle+i\langle k_T^i\frac{\partial K}{\partial T^i}\rangle+i\langle k_Z\frac{\partial K}{\partial Z}\rangle \nonumber \\
&= -\frac{a\epsilon_S\epsilon_R^2}{2\kappa^2_4V^{2/3}\hat R}\left( \mu(L) +\frac{\pi\epsilon_S \hat R}{V^{1/3}}
\left(\beta_i^{(2)} +  W_iz^2  \right)  l^i  \right)\ ,
\label{rain5}
\end{align}
that is, the part of  $\langle \mathcal{P}  \rangle$ that is independent of the matter field VEVs $\langle C^{L} \rangle$. Note that the moduli dependent parameter from the expression for the FI term, such as $V$ and $\hat R$, have fixed values inside the vacuum we have just described. The reader should be aware that we \emph{dropped} the VEV bracket notation for brevity. Expression \eqref{rain4} can now simply be written as
\begin{equation}
FI+i \langle k_C^L\frac{\partial K}{\partial C^L}\rangle =0 \ .
\label{rain4A}
\end{equation}
The value of the FI term, whether it is vanishing or non-vanishing, will play an important role below in categorizing specific vacua. However, for the remainder of this section we will leave the FI term unspecified.

To continue, let us expand the moduli and matter scalar fields as 
\begin{equation}
\label{hope1}
S=\langle S\rangle +\delta S\ , \quad T^i=\langle T^i\rangle +\delta T^i\ ,\quad Z=\langle Z\rangle +\delta Z\ ,\quad C^L=\langle C^L\rangle +\delta C^L\ ,
\end{equation}
where the VEVs $\langle S \rangle$, $\langle T^{i} \rangle$, $\langle Z \rangle$ and $\langle C^{L} \rangle$ satisfy the $D$-flatness condition \eqref{rain4}.
Inserting this into Lagrangian \eqref{rain1}, one can compute the effective Lagrangian involving the scalar fluctuations $\delta S,\>\delta T^i$, $\delta Z$ and $\delta C^L$, as well as their fermionic superpartners. This process is discussed in detail in Appendix B. Here, we will simply present the results.

\subsection{Massive U(1) Vector Superfield}

We begin by considering the purely scalar field part of Lagrangian $\cal{L}$ given in \eqref{rain1}; that is
\begin{equation}
\begin{split}
\label{hope2}
&- g_{S\bar S}D_\mu S D^\mu \bar S-g_{T^i\bar T^j}D_\mu T^i D^\mu \bar T^{\bar j}-g_{Z\bar Z}D_\mu Z D^\mu \bar Z- g_{C^L\bar C^{\bar M}}D_\mu C^L D^\mu \bar C^{\bar M}\\
&(- g_{S\bar T^i}D_\mu S D^\mu \bar T^i- g_{S\bar Z}D_\mu S D^\mu \bar Z- g_{T^i\bar C^L}D_\mu T^i D^\mu \bar C^L
 - g_{Z\bar T^i}D_\mu Z D^\mu \bar T^i+hc) -\tfrac{g_2^2}{2}\mathcal{P}^2 \ .
 \end{split}
\end{equation}
Expanding the scalar fields around their VEVs as in \eqref{hope1}, it is clear that the kinetic energy terms for $\delta S, \delta T^{i}$, $\delta Z$ and $\delta C^{L}$ are not canonically normalized. Following the notation of Appendix B, let us 
write
\begin{equation}
\delta z^{A}, A=1,\dots,N ~\equiv~ \delta S, \delta T^{i},\delta Z,\delta C^{L} \ , \quad N=2+h^{1,1}+\cal{N} 
\label{hope3}
\end{equation}
and define new complex scalar fields $\xi^{A}, A=1,\dots,N$ as linear combinations of the $\delta z^{A}$
\begin{equation}
\label{hope4}
\xi^{A}=[U^{A}_{B}] \delta z^{B} \ .
\end{equation}
As shown in Appendix B, the matrix $[U^{A}_{B}]$ can indeed by chosen so that the kinetic terms for all of the $\xi^{A}$ fields are canonically normalized. However, this matrix is not unique. Be that as it may, it is convenient to specify the first row of this matrix such that
\begin{equation}
\label{hope5}
\xi^{1}=\frac{ \langle g_{A \bar{B}} \bar{k}^{\bar{B}} \rangle} {  \sqrt{ \langle  g_{C \bar{D}}k^{C} \bar{k}^{\bar{D}} \rangle }   } \delta z^{A} \ .
\end{equation}
Using \eqref{rain2} and \eqref{black1}, it is straight forward to show that the kinetic term for $\xi^{1}$ is $-\partial_{\mu} \xi^{1} \partial ^{\mu} \bar{\xi}^{1}$ and, hence, is canonically normalized. Complex field $\xi^{1}$ defined in \eqref{hope5} is chosen for the following two reasons. 

To begin with, let us consider the potential energy $V_{D}=\frac{1}{2}g_2\mathcal{P}^2$ defined in \eqref{black2}. Expanding the scalar fields as in \eqref{hope1} and choosing their VEVs such that $ \langle \mathcal{P} \rangle = 0 $, it follows that
\begin{equation}
\mathcal{L} \supset -\frac{1}{2}\langle g_2^{2}\rangle {\delta \mathcal{P}}^{2} = -2 \langle g_{2}^{2}g_{A\bar{B}}k^{A} \bar{k}^{\bar{B}} \rangle \phi^{2} \ ,
\label{black3}
\end{equation}
where scalar field $\phi$ is given by
\begin{equation}
\phi=\frac{ i\langle k^Ag_{A\bar B}\rangle \tfrac{\delta \bar z^{\bar B}}{2}-i\langle k^{\bar B}g_{A\bar B}\rangle \tfrac{\delta z^A}{2}  }{\sqrt{ \langle g_{A\bar B}k^A\bar k^{\bar B}\rangle}}\ .\\
\label{black4}
\end{equation}
Secondly, writing out the part of the kinetic terms in \eqref{hope2} that, because of the covariant derivatives, involve the gauge connection either linearly as $A_{2\mu}$ or quadratically as $A_{2\mu}A_{2}^{\mu}$, one finds that
\begin{equation}
\mathcal{L} \supset 2 \sqrt{\langle g_{A \bar{B}}k^{A}\bar{k}^{\bar{B}} \rangle} \partial_{\mu} \eta A_{2}^{\mu} - \langle g_{A \bar{B}}k^{A}\bar{k}^{\bar{B}} \rangle A_{2}^{\mu}A_{2 \mu} \ ,
\label{black5}
\end{equation}
where $\eta$ is a real scalar field given by
\begin{equation}
\label{black6}
\eta=\frac{ \langle g_{A\bar{B}} k^{A} \rangle \frac{\delta \bar{z}^{\bar{B}}}{2}+  \langle g_{A\bar{B}}\bar{k}^{\bar{B}} \rangle \frac{\delta {z}^{A}}{2}    }{ \sqrt{ \langle  g_{C \bar{D}}k^{C} \bar{k}^{\bar{D}} \rangle }  } \ .
\end{equation}
The role of $\eta$ is more subtle, but very important--as we will demonstrate momentarily. 

Having defined and motivated the two real fields $\phi$ and $\eta$, we can now explain our choice of \eqref{hope5} for the complex field $\xi^{1}$. It is simply that
\begin{equation}
\xi^{1}= \eta+i\phi \ .
\label{black7}
\end{equation}
Using the fact that the kinetic term for $\xi^{1}$ is canonically normalized to unity, it follows that 
\begin{equation}
\label{black8}
-\partial_{\mu}\xi^{1}\partial^{\mu} \bar{\xi}^{1}= -\partial_{\mu} \phi \partial^{\mu} \phi -\partial_{\mu} \eta \partial^{\mu} \eta \ .
\end{equation}
Combining the first term \eqref{black8} with the $D$-term potential \eqref{black3}, it follows that
\begin{equation}
\label{black9}
\mathcal{L} \supset -\partial_{\mu} \phi \partial^{\mu} \phi  -2 \langle g_{2}^{2} g_{A\bar{B}}k^{A} \bar{k}^{\bar{B}} \rangle\phi^{2}
\end{equation}
and, hence, $\phi$ is a real scalar field arising in the Lagrangian with mass
\begin{equation}
\label{black10}
m_{\phi}= \sqrt{2\langle g_{2}^{2} g_{A\bar{B}}k^{A}\bar{k}^{\bar{B}}\rangle} \ .
\end{equation}
Next, combining the second term in \eqref{black8} with \eqref{black5}, it follows that 
\begin{equation}
\label{black11}
\mathcal{L} \supset  -\langle g_{A \bar{B}}k^{A} \bar{k}^{\bar{B}}\rangle  \Big( A_{2\mu}-\frac{\partial_{\mu}\eta}{ \sqrt{\langle g_{A \bar{B}}k^{A} \bar{k}^{\bar{B}}\rangle}} \Big) \Big(A_{2}^{\mu} -\frac{\partial^{\mu}
\eta}{ \sqrt{\langle g_{A \bar{B}}k^{A} \bar{k}^{\bar{B}}\rangle}} \Big)\ .
\end{equation}
Redefining the field to $A_{2\mu}^{\prime}$ via the gauge transformation
\begin{equation}
\label{black12}
A_{2\mu}^{\prime}=A_{2\mu}-\frac{\partial_{\mu}\eta}{ \sqrt{\langle g_{A \bar{B}}k^{A} \bar{k}^{\bar{B}}\rangle}} \ ,
\end{equation}
it follows that, adding the gauge kinetic term in \eqref{rain1},  \eqref{black11} simply becomes
\begin{equation}
\label{black13}
-\tfrac{1}{4g^2_2}F^{\mu \nu}_2 F_{2\mu \nu}-\langle g_{A \bar{B}}k^{A}\bar{k}^{\bar{B}} \rangle A_{2\mu}A_{2}^{\mu}\ ,
\end{equation}
where we have dropped the prime on $A_{2\mu}$. That is, the scalar field $\eta$ is simply the Goldstone boson which is gauged away and disappears from the Lagrangian. As shown in Appendix B, the gauge field then acquires a mass
given by
\begin{equation}
\label{black14}
m_{A}=\sqrt{2 \langle g_{2}^{2} g_{A\bar{B}}k^{A} \bar{k}^{\bar{B}} \rangle} \ .
\end{equation}
Note that this is identical to the scalar mass presented in \eqref{black10}.

Let us now consider the fermion quadratic terms in Lagrangian \eqref{rain1}. These are
\begin{equation}
\begin{split}
& -ig_{S\bar S}\psi_S \slashed{\mathcal{D}} \psi_S^{\dag}
 -ig_{T^i\bar T^j}\psi_T^i \slashed{\mathcal{D}} \psi_T^{\bar j\dag}  -ig_{Z\bar Z}\psi_Z \slashed{\mathcal{D}} \psi_Z^{\dag}-ig_{C^L\bar C^{\bar M}}\psi^L \slashed{\mathcal{D}} \psi^{\bar M\dag}+\cdots\\
&+\sqrt{2}\Big[g_{S\bar S}k_S\lambda_{2}^\dag \psi_S+g_{S\bar S}\bar k_S\lambda_{2} \psi_S+g_{T^i\bar T^j}k_T^i\lambda_{2}^\dag \psi_T^{\bar j\dag}+g_{T^i\bar T^j}\bar k_T^{\bar j}\lambda_{2} \psi_T^i+g_{Z\bar Z}k_Z\lambda_{2}^\dag \psi_Z\\&+g_{Z\bar Z}\bar k_Z\lambda_{2} \psi_Z+g_{C^L\bar C^{\bar M}}k_C^L\lambda_{2}^\dag \psi^{\bar M\dag}+g_{C^L\bar C^{\bar M}}\bar k_C^{\bar M}\lambda_{2} \psi_C^L+\dots\Big]-\frac{i}{g_2^2} \lambda_{2} \slashed{\partial} \lambda_{2}^{\dag}\ . 
\label{repair1}
\end{split}
\end{equation}

As discussed in Appendix B, the fermions are rotated into the physical, canonically normalized states using the same $[U^{A}_{B}]$ matrix as for the scalar fields; that is
\begin{equation}
\label{repair2}
\psi_{\xi}^{A}=[U^{A}_{B}] \psi^{B} \ .
\end{equation}
As mentioned above, this matrix is not completely specified with the exception of the first row. It follows from \eqref{hope5} that
\begin{equation}
\label{repair3}
\psi_{\xi}^{1}=\frac{ \langle g_{A \bar{B}} \bar{k}^{\bar{B}} \rangle} {  \sqrt{ \langle  g_{C \bar{D}}k^{C} \bar{k}^{\bar{D}} \rangle }   } \psi^{A} \ .
\end{equation}
As with the scalar $\xi^{1}$, let us now examine the terms in the Lagrangian involving the canonically normalized fermion $\psi_{\xi}^{1}$. We find that, adding the gaugino kinetic term in \eqref{rain1}, 
\begin{equation}
\mathcal{L} \supset  -i\psi_{\xi}^{1}\slashed{\partial} \psi_{\xi}^{1\dagger}-\frac{i}{g_2^2}\lambda_{2} \slashed{\partial} \lambda_{2}^\dag + \sqrt{2\langle g_{2}^{2} g_{A\bar{B}}k^{A}\bar{k}^{\bar{B}}\rangle} \big(\psi_{\xi}^{1\dagger}\frac{\lambda_{2}^\dag}{\langle g_2\rangle} + \psi_{\xi}^{1} \frac{\lambda_{2}}{\langle g_2\rangle}  \big) \ .
\label{repair4}
\end{equation}
Defining the Dirac spinor
\begin{equation}
\label{repair5}
\Psi= \dbinom{\lambda_{2}^{\dagger}/\langle g_2\rangle}{\psi_{\xi}^{1}} \ ,
\end{equation}
\eqref{repair4} can be written as
\begin{equation}
\mathcal{L} \supset \bar{\Psi} \slashed{\partial} \Psi + \sqrt{2\langle g_{2}^{2} g_{A\bar{B}}k^{A}\bar{k}^{\bar{B}}\rangle} \bar{\Psi}{\Psi} \ .
\label{repair6}
\end{equation}
Hence, $\Psi$ is a Dirac spinor in the Lagrangian with mass
\begin{equation}
\label{repair7}
m_{\Psi}= \sqrt{2\langle g_{2}^{2} g_{A\bar{B}}k^{A}\bar{k}^{\bar{B}}\rangle} \ .
\end{equation}
Note that this is identical to the scalar and gauge connection masses given in \eqref{black10} and \eqref{black14} respectively. 

Combining all of the results of this Section, we conclude that the scalar field $\phi$, the massive gauge field $A_{2\mu}$ and the Dirac spinor $\Psi$ together form the component fields of a massive Abelian vector superfield of the form
\begin{equation}
\label{repair8}
(\phi, A_{2\mu}, \Psi )
\end{equation}
with mass
\begin{equation}
\label{repair9}
m_{\phi}=m_{A}=m_{\Psi}= \sqrt{2\langle g_{2}^{2} g_{A\bar{B}}k^{A}\bar{k}^{\bar{B}}\rangle} \ .
\end{equation}
Using the expressions for $g_{A \bar{B}}$ and $k^{A}$ given in \eqref{black1} and \eqref{rain2} respectively, we find that the squared mass of this vector superfield can be expressed as
\begin{equation}
m_{A}^{2}=m_{\text{anom}}^{2}+m_{\text{matter}}^{2} \ ,
\label{repair10}
\end{equation}
where 
\begin{equation}
\begin{split}
m_{\text{anom}}^{2} =\frac{\pi \hat{\alpha}_{\text{GUT}}}{a\text{Re} f_{2} }\>8a^{2}\epsilon_{S}^{2}\epsilon_{R}^{4} \Big[& g^{T}_{i \bar{j}}l^{i}l^{\bar{j}} +\frac{\pi^2\epsilon_S^2}{8V^2}\left(  \beta^{(2)}_il^i-
W_il^iz^2 \right)^2\Big]
\label{repair11}
\end{split}
\end{equation}
is the ``anomalous'' mass due to the inhomogeneous transformation of the $S$ and $T^{i}$ axions and
\begin{equation}
\begin{split}
\label{repair12}
m_{\text{matter}}^{2} &=\frac{\pi \hat{\alpha}_{\text{GUT}}}{a\text{Re}  f_{2} } 2   e^{\kappa_4^2 K_T/3}\mathcal{G}_{L\bar{M}}  \langle C^{L} \rangle \langle \bar C^{\bar{M}} \rangle\left( Q^LQ^{\bar M}+\frac{1}{3}a\epsilon_S\epsilon_R^2Q^{\bar M}\frac{\mu(L)}{\hat RV^{2/3}} \right)\\
&+\frac{\pi \hat{\alpha}_{\text{GUT}}}{a\text{Re}  f_{2} } 8a^{2}\epsilon_{S}^{2}\epsilon_{R}^{4}\left(\kappa_4^2g_{ij}^Tl^il^j+\frac{1}{3}\left(\frac{\mu(L)}{4\hat RV^{2/3}}\right)^2\right)\frac{e^{\kappa_4^2K_T/3}}{3}\mathcal{G}_{L\bar M}\langle C^{L} \rangle \langle \bar C^{\bar{M}} \rangle\\
&=\frac{\pi \hat{\alpha}_{\text{GUT}}}{a\text{Re}  f_{2} } 2  {G^\prime}_{L\bar{M}}  \langle C^{L} \rangle \langle \bar C^{\bar{M}} \rangle
\end{split}
\end{equation}
is the usual mass generated by the Higgs mechanism associated with non-zero VEVs of the matter scalars $C^{L}$. Note that in the above expression we have defined yet another metric,
\begin{equation}
\begin{split}
\label{def_GLMp}
G_{L\bar M}^\prime=  e^{\kappa_4^2 K_T/3}\mathcal{G}_{L\bar{M}} & \Bigg[ Q^LQ^{\bar M}+\frac{1}{3}\frac{a\epsilon_S\epsilon_R^2}{\hat RV^{2/3}}\mu(L) 
\\&+ \frac{4}{3}{a^{2}\epsilon_{S}^{2}\epsilon_{R}^{4}}\left(\kappa_4^2g_{ij}^Tl^il^j+\frac{1}{3}\left(\frac{\mu(L)}{4\hat RV^{2/3}}\right)^2\right)\Bigg]\ ,
\end{split}
\end{equation}
which is different that the metric $G_{L\bar M}$ defined in eq. \eqref{def_GLM}.
Once again, we have, for simplicity, dropped the VEV bracket notation for the moduli dependent parameters, such as $V$, $\hat R$, and $f_2$, in the above expressions.

We conclude that the chiral superfield with components $(\xi^{1}, \psi_{\xi}^{1})$ becomes part of the massive vector multiplet and is no longer in the light effective Lagrangian. 

\subsection{Low Mass Effective Lagrangian}

For any valid choice of the rotation matrix $[U^{A}_{B}]$, there remain $A=2,\dots,N$ massless chiral superfields  with components $(\xi^{A}, \psi^{A}_{\xi})$. Their low energy Lagrangian is simply given by
\begin{equation}
\mathcal{L} \supset \sum_{A=2}^{N} \Big(-\partial_{\mu}\xi^{A}\partial^{\mu} \bar{\xi}^{\bar{A}} -i\psi_{\xi}^{A}\slashed{\partial} \psi_{\xi}^{\bar{A}\dagger} \Big) \ .
\label{repair13}
\end{equation}
We will discuss this massless low energy Lagrangian, extended to include interaction terms, in future publications. 

The content of this Section is rather generic, allowing for an arbitrary number of K\"ahler moduli and matter scalars $C^L$. It is meant as a more descriptive discussion of Appendix B. However, it is useful to give a specific example for a vacuum with a small number of K\"ahler moduli and a calculable set of matter fields, as well as a specific choice of the FI term. We will do this in the next two Sections.

\section{A Specific Hidden Sector with FI=0}

To begin, let us consider CY threefolds with three K\"ahler moduli; that is, we henceforth choose $h^{1,1}=3$. Therefore, the chiral superfields that appear in this Section are $\tilde{S}$, $\tilde{T}^{i}, i=1,2,3$, $\tilde Z$ and, for the time being, an arbitrary number of matter chiral superfields $\tilde{C}^{L}, L=1,\dots, \cal{N}$. We say ``for the time being'' because, given an explicit hidden sector vacuum with $h^{1,1}=3$, the number of matter superfields can be explicitly calculated. For example, in the hidden sector of the $B-L$ MSSM vacuum presented in~\cite{Ashmore:2020ocb} and ~\cite{Ashmore:2020wwv}, which indeed has $h^{1,1}=3$, we showed that ${\cal{N}}=58$.

Second, recall that for the hidden sector vacuum to be $N=1$ supersymmetric, it is necessary for the vacuum expectation values of the scalar fields $S$, $T^i$, $Z$ and $C^L$ to satisfy the D-flatness condition
\begin{equation}
V=V_D=0\quad \Rightarrow \quad \langle \mathcal{P}  \rangle =0
\end{equation} 
or, equivalently, from \eqref{rain4A} that
\begin{equation}
FI+i \langle k_C^L\frac{\partial K}{\partial C^L}\rangle =0 \ ,
\label{x1}
\end{equation}
where FI is the Fayet-Iliopoulos term defined in \eqref{rain5} by
\begin{equation}
FI=i\langle k_S\frac{\partial K}{\partial S}\rangle+i\langle k_T^i\frac{\partial K}{\partial T^i}\rangle+i\langle k_Z\frac{\partial K}{\partial Z}\rangle  \ .
\label{x2}
\end{equation}
In this Section, we will explicitly choose the VEVs $\langle S \rangle$, $\langle T^{i} \rangle, i=1,2,3$ and $\langle Z\rangle$ so that the Fayet-Iliopoulos term vanishes. It then follows from \eqref{rain5} that 
\begin{equation}
FI= -\frac{a\epsilon_S\epsilon_R^2}{2\kappa^2_4V^{2/3}\hat R}\left( \mu(L) +\frac{\pi\epsilon_S \hat R}{V^{1/3}}
\left(\beta_i^{(2)} +  W_iz^2  \right)  l^i  \right) = 0 \ .
\label{x3}
\end{equation}
Note from \eqref{rain3} that for vacua with a vanishing FI term
\begin{equation}
\langle C^{L} \rangle = 0, \quad L=1, \dots, \cal{N} \ .
\label{x4}
\end{equation}
It then follows from \eqref{rain2} that 
\begin{equation}
\label{x4}
 \langle k_C^L \rangle=-iQ^L \langle C^L \rangle =0, \quad L=1, \dots, \cal{N}
\end{equation}
and, hence, many of the generic expressions in Section 2 greatly simplify. Henceforth, in this Section, the VEVs of the scalar fields satisfy
\begin{equation}
\begin{split}
&\langle S\rangle\neq 0\ , \langle T^i\rangle\neq 0\ ,\langle Z\rangle \neq 0\ ,\quad i=1,2,3\ ,\\
&\langle C_L\rangle =0\ ,\quad L=1,\dots, \mathcal{N}\ .
\end{split}
\label{x5}
\end{equation}

When the FI term vanishes, the D-term stabilization mechanism fixes a combination of the real parts of the dilaton, the K\"ahler moduli and the five-brane modulus only. The D-flatness condition determines a so-called ``vanishing slope surface'' in the moduli space--see~\cite{Ashmore:2020ocb} for a detailed account. The moduli can take values within this subspace only. Equivalently, one linear combination of the moduli has been fixed and in reality, the effective has one less modulus. Fixing the VEVs of remaining moduli would require an additional moduli stabilization mechanism~\cite{Conlon:2010ji,Anderson:2011cza}, which however, is beyond the scope of this work.
The matter superfields $C^{L}$ have vanishing scalar VEVs but are otherwise independent. For example, since $\langle k_{C}^{L} \rangle=0$, the mixing terms of the matter fermions $\psi_{C}^{L}$ with the $U(1)$ gaugino $\lambda_{2}$ in Lagrangian \eqref{rain1} vanish. Hence, to canonically normalize  the chiral superfields and diagonalize their mass matrix, one need only consider the superfields $\tilde{S}$, $\tilde{T}^{i}, i=1,2,3$ and $\tilde Z$. The matter chiral superfields $\tilde{C}^{L}$ decouple and can be treated independently. Therefore, we will ignore them until the end of this Section. That is, to obtain the canonically normalized low energy spectrum, we use the results of Appendix B with $z^A=(S,T^1,T^2,T^3,Z)$ only. Following Appendix B, we perturb the moduli fields around their VEVs, such that
\begin{equation}
S=\langle S\rangle +\delta S\ , \quad T^i=\langle T^i\rangle +\delta T^i\ ,\quad Z=\langle Z\rangle +\delta Z\ .
\end{equation}
We then rotate the basis composed of these scalar perturbations,
as well as their fermionic superpartners, $\psi_S$, $\psi_T^{i}$ and $\psi_Z$ respectively, into a new basis $(\xi^A ,\psi_\xi^A)$, for $A=1,\dots,2+h^{1,1}=5$. As shown in Appendix B, the change of basis is achieved via a rotation matrix $[U_{B}^{A}]$, which is the same for both the scalars and the fermions, such that 
\begin{equation}
\begin{split}
&\xi^A=U^A_1\delta S+\sum_{i=1}^3U^A_i\delta T^i+U_5^A\delta Z\ ,\quad A=1,\dots,5, \quad \\
&\psi_\xi^A=U^A_1\psi_S+\sum_{i=1}^3U^A_i\psi_T^i+U_5^A\psi_Z\ .
\end{split}
\end{equation}
%
%
The form of the rotation matrix $[U_{A}^{B}]$ is given by \eqref{eq:matrix_solada} in Appendix B. For the specific vacua of this Section, consisting of one dilaton, three K\"ahler moduli fields and one five-brane modulus, $[U_{A}^{B}]$ is a $5 \times 5$ matrix.
As discussed in Appendix B, only the first row of this matrix is determined. The elements of this row have the form
{\small
\begin{equation}
\begin{split}
\label{first_row}
&\left(U^1_1\ ,U^1_2\ ,U^1_3\ ,U^1_4\ ,U^1_5\right)\equiv\\
&\left(\frac{\langle g_{S\bar B}\bar k^{\bar B}\rangle}{\sqrt{\langle g_{B\bar C}k^B\bar k^{\bar C}\rangle}}\ ,
\frac{\langle g_{T^1\bar B}\bar k^{\bar B}\rangle}{\sqrt{\langle g_{B\bar C}k^B\bar k^{\bar C}\rangle}}\ ,
\frac{\langle g_{T^2\bar B}\bar k^{\bar B}\rangle}{\sqrt{\langle g_{B\bar C}k^B\bar k^{\bar C}\rangle}}\ ,
\frac{\langle g_{T^3\bar B}\bar k^{\bar B}\rangle}{\sqrt{\langle g_{B\bar C}k^B\bar k^{\bar C}\rangle}}\ ,
\frac{\langle g_{Z\bar B}\bar k^{\bar B}\rangle}{\sqrt{\langle g_{B\bar C}k^B\bar k^{\bar C}\rangle}}\right)\ ,
\end{split}
\end{equation}
}
\noindent where the indices, $B,C=1,\dots, 5$ run over the moduli $S, T^1, T^2,T^3,Z$. In this notation, 
\begin{equation}
 k^{\bar B}=( k_S,\> k_T^1,\> k_T^2,\> k_T^3,\> k_Z), \quad \bar k^{\bar B}=(\bar k_S,\>\bar k_T^1,\>\bar k_T^2,\>\bar k_T^3,\>\bar k_Z)\ .
\end{equation}
For specificity, we compute these matrix elements  using the expressions for the Kahler metrics, derived in Appendix C, and the expressions for the Killing vectors $k_S,\>k_T^1,\>k_T^2,\>k_T^3$ and $k_Z$ from \eqref{rain2}. We find that
{\small
\begingroup
\allowdisplaybreaks
\begin{align}
\nonumber U_1^1&=
\frac{g_{S\bar S}\bar k_{ S}+\sum_{i=j}^3g_{S\bar T^j}\bar k^{ j}_T+g_{S\bar Z}\bar k_{ Z}}{\sqrt{\langle g_{A\bar B}k^A\bar k^{\bar B}\rangle}}=\frac{\tfrac{1}{8V^2}\pi\epsilon_S(\beta_i^{(2)}+W_iz^2)l^i}{\sqrt{ \kappa_4^2g^{T}_{i \bar{j}}l^{i}l^{\bar{j}} +\frac{\pi^2\epsilon_S^2}{8V^2}\left(  \beta^{(2)}_il^i-
W_il^i z^2\right)^2}}
 \ ,\\
\nonumber U_i^1&=
\frac{g_{T^i\bar S}\bar k_{ S}+\sum_{j=1}^3g_{T^i\bar T^j}\bar k^{ j}_T+g_{T^i\bar Z}\bar k_{ Z}}{\sqrt{\langle g_{A\bar B}k^A\bar k^{\bar B}\rangle}}=\frac{\kappa_4^2g^T_{ij}l^j+\tfrac{1}{16V^2}\pi^2\epsilon^2_SW_iz^2(\beta_j^{(2)}+W_jz^2)l^j}{\sqrt{ \kappa_4^2g^{T}_{i \bar{j}}l^{i}l^{\bar{j}} +\frac{\pi^2\epsilon_S^2}{8V^2}\left(  \beta^{(2)}_il^i-
W_il^i z^2\right)^2}}\ ,\\
U_5^1&=
\frac{g_{Z\bar S}\bar k_{ S}+\sum_{i=j}^3g_{Z\bar T^j}\bar k^{ j}_T+g_{Z\bar Z}\bar k_{ Z} }{\sqrt{\langle g_{A\bar B}k^A\bar k^{\bar B}\rangle}}=-\frac{\tfrac{1}{8V^2}{\pi^2\epsilon_S^2z(\beta_i^{(2)}+W_iz^2)l^i}}{\sqrt{ \kappa_4^2g^{T}_{i \bar{j}}l^{i}l^{\bar{j}} +\frac{\pi^2\epsilon_S^2}{8V^2}\left(  \beta^{(2)}_il^i-
W_il^iz^2 \right)^2}}\ ,
\end{align}
\endgroup
}
where $i=1,2,3$. The remaining four rows contain 20 complex coefficients $u_{2,3,4,5}^{S,T^{i},Z}$ that are not specified. As we will discuss in detail below, these coefficients are subject to a number of non-trivial constraints which will determine all but six of them. For the time being, however, let us consider 
only the component fields in the supermultiplet $(\xi^1,\psi^1_{\xi})$--since they are are completely specified given the matrix elements shown above. As discussed in Section 2, complex scalar field $\xi^1$ naturally decomposes as
\begin{equation}
\xi^{1}=\eta+i\phi
\label{gd3}
\end{equation}
Using expressions \eqref{black6} and \eqref{black4} respectively--as well as the Killing fields $k_{S}$, $k_{T}^{i}$, $k_Z$ given in \eqref{rain2} and the moduli space metrics presented in Appendix C, now restricted to $i,j=1,2,3$--the scalar field components are given, to linear order in $\epsilon_S$, by
\begin{equation}
\eta=\frac{ \frac{1}{8V^2}\pi \epsilon_S\left( \beta^{(2)}_i+  W_iz^2\right) l^i\sigma +2\kappa_4^2g^T_{i\bar j}l^{\bar j}\chi^{i}}
{\sqrt{\kappa_4^2g^{T}_{i \bar{j}}l^{i}l^{\bar{j}} }}+\mathcal{O}(\epsilon_S^2) \ ,
\end{equation}
which forms the longitudinal degree of freedom of the massive vector boson $A_{2\mu}$, and 
\begin{equation}
 \phi=\frac{ \frac{1}{8V^2}\pi \epsilon_S\left( \beta^{(2)}_i+  W_iz^2\right) l^i\delta V +\kappa_4^2g^T_{i\bar j}l^{\bar j}\delta t^i}
{\sqrt{\kappa_4^2g^{T}_{i \bar{j}}l^{i}l^{\bar{j}} } }+\mathcal{O}(\epsilon_S^2)\ ,
\end{equation}
which is the scalar component of the massive gauge vector supermultiplet. Note that, for simplicity,  we have \emph{dropped the VEV bracket notation} from these final expressions for all moduli dependent quantities. We will continue to do this in every {\it final} result throughout the rest of this section. Note that at order $\epsilon_S^2$, the five brane axion and the five-brane scalar perturbation $\delta z$ enter the definitions of $\phi$ and $\eta$ as well.

Similarly, the fermion $\psi^1_{\xi}$ is a linear combination of the fermionic parts of the chiral moduli multiplets. Using the expressions for the Killing fields $k_{S}$, $k_{T}^{i}$, $k_Z$ given in \eqref{rain2} and for the moduli space metrics presented in Appendix C, now restricted to $i,j=1,2,3$, we find that
\begin{equation}
 \psi_\xi^1=\frac{ \frac{1}{4V^2}\pi \epsilon_S\left( \beta^{(2)}_i+  W_iz^2\right) l^i\psi_S +\kappa_4^2g^T_{i\bar j}l^{\bar j}\psi_T^i}
{\sqrt{\kappa_4^2g^{T}_{i \bar{j}}l^{i}l^{\bar{j}}} }+\mathcal{O}(\epsilon_S^2)\ .
\end{equation}
As discussed in Section 2,  $\psi^1_{\xi}$ combines with the gaugino $\lambda_{2}$ to form a massive Dirac fermion
\begin{equation}
\label{late1}
\Psi= \dbinom{\lambda_{2}^{\dagger}/\langle g_2\rangle}{\psi_{\xi}^{1}} \ .
\end{equation}

As discussed in Section 2, the masses of the fields $\phi$, $ A_{2\mu}$, and $ \Psi $ are all identical and, hence, they form the components of a massive vector superfield
\begin{equation}
\label{late2}
(\phi, A_{2\mu}, \Psi )
\end{equation}
whose mass is given in \eqref{repair9}. 
Using the expressions for $g_{A \bar{B}}$ and $k^{A}$ given in Appendix C and \eqref{rain2} respectively, we find that the squared mass of this vector superfield can be expressed as
\begin{equation}
m_{A}^{2}=m_{\text{anom}}^{2}+m_{\text{matter}}^{2} \ ,
\label{repair10A}
\end{equation}
where it follows from \eqref{repair11} that
\begin{equation}
\begin{split}
m_{\text{anom}}^{2} =\frac{\pi \hat{\alpha}_{\text{GUT}}}{a\text{Re} f_{2} }\>8a^{2}\epsilon_{S}^{2}\epsilon_{R}^{4} \Big[& g^{T}_{i \bar{j}}l^{i}l^{\bar{j}} +\frac{\pi^2\epsilon_S^2}{8V^2}\left(  \beta^{(2)}_il^i-
W_il^i z^2\right)^2\Big]
\label{late4}
\end{split}
\end{equation}
is the ``anomalous'' mass due to the inhomogeneous transformation of the $S$ and $T^{i}$ axions, and from \eqref{repair12} that
\begin{equation}
m_{\text{matter}}^{2} = 0 \ ,
\end{equation}
since for vacua with FI=0,  the VEVs of the matter fields $C^{L}$ all vanish. That is, for vacua for which FI=0, the mass of the vector supermultiplet \eqref{late2} is due entirely to the inhomogeneous anomalous $U(1)$ transformations of the $S$ and $T^{i}$ axions,

%
%
%

In Appendix B, we have shown that the row vectors of the matrix $[U_{A}^{B}]$ for $A=2,3,4,5$ can be written in the form
{\small
\begin{equation}
\begin{split}
\label{TS_row_vect}
\underline{g\cdot u_A}&={\Bigg(\frac{\langle g_{S\bar B}\rangle \bar u_A^B}{\sqrt{ \langle g_{B\bar C}u_A^B\bar u^{\bar C}_A\rangle}} ,\>\frac{\langle g_{T^1\bar B}\rangle \bar u_A^B}{\sqrt{ \langle g_{B\bar C}u_A^B\bar u^{\bar C}_A\rangle}} ,\> \frac{\langle g_{T^2\bar B}\rangle \bar u_A^B}{\sqrt{ \langle g_{B\bar C}u_A^B\bar u^{\bar C}_A\rangle}} ,\>\frac{\langle g_{T^3\bar B}\rangle \bar u_A^B}{\sqrt{ \langle g_{B\bar C}u_A^B\bar u^{\bar C}_A\rangle}},\>\frac{\langle g_{Z\bar B}\rangle \bar u_A^B}{\sqrt{ \langle g_{B\bar C}u_A^B\bar u^{\bar C}_A\rangle}}  \Bigg)} 
\end{split}
\end{equation}
}
where the indices $B,C=1,\dots, 5$ run over the moduli $S, T^1, T^2,T^3,Z$ and there are 20 complex coefficients $u_{2,3,4,5}^{S,T^{i},Z}$  which cannot, a priori, be determined. 
It follows that
the rest of the canonically normalized states, namely 
\begin{equation}
\begin{split}
&\xi^A=\tfrac{\langle g_{S\bar B}\rangle \bar u_A^B}{\sqrt{ \langle g_{B\bar C}u_A^B\bar u^{\bar C}_A\rangle}}\delta S+\sum_{i=1}^3 \tfrac{\langle g_{T^i\bar B}\rangle \bar u_A^B}{\sqrt{ \langle g_{B\bar C}u_A^B\bar u^{\bar C}_A\rangle}}\delta T^i+\tfrac{\langle g_{Z\bar B}\rangle \bar u_A^B}{\sqrt{ \langle g_{B\bar C}u_A^B\bar u^{\bar C}_A\rangle}}\delta Z\ ,
\end{split}
\end{equation}
for $A=2,3,4,5$, form massless chiral supermultiplets together with their femionic superpartners, 
\begin{equation}
\begin{split}
&\psi_\xi^A=\tfrac{\langle g_{S\bar B}\rangle \bar u_A^B}{\sqrt{ \langle g_{B\bar C}u_A^B\bar u^{\bar C}_A\rangle}} \psi_S+\sum_{i=1}^3 \tfrac{\langle g_{T^i\bar B}\rangle \bar u_A^B}{\sqrt{ \langle g_{B\bar C}u_A^B\bar u^{\bar C}_A\rangle}}\psi_T^i+\tfrac{\langle g_{Z\bar B}\rangle \bar u_A^B}{\sqrt{ \langle g_{B\bar C}u_A^B\bar u^{\bar C}_A\rangle}}\psi_Z\ .
\end{split}
\end{equation}
%

In the absence of an explicit moduli stabilization mechanism to fix all of the moduli VEVs to specific values, these states, which are orthogonal to the fields $\xi^1$ and $\psi^1_{\xi}$, represent flat directions in the theory with K\"ahler potential
\begin{equation}
K^{\xi}=\xi^2\xi^{2\dagger}+\xi^{3}\xi^{3\dagger}+\xi^4\xi^{4\dagger}+\xi^5\xi^{5\dagger}  \ .
\end{equation}
The 20 complex coefficients $u_{2,3,4,5}^{S,T^{i},Z}$ cannot be determined in the absence of an additional potential which can completely specify the VEVs of all the moduli fields. However, these 20 vector parameters are not independent. 
%
The row vectors presented in \eqref{TS_row_vect} define the column vectors of the inverse matrix $[U_{A}^{B}]^{-1}$ given by 
\begin{equation}
\label{TS_column_vect}
\underline{u_{A}}=
\left(
\begin{matrix}
\frac{u_A^{S}}{\sqrt{ \langle g_{B\bar C}u_A^B\bar u^{\bar C}_A\rangle}}\\
\frac{u_A^{T^1}}{\sqrt{ \langle g_{B\bar C}u_A^B\bar u^{\bar C}_A\rangle}}\\
\frac{u_A^{T^2}}{\sqrt{ \langle g_{B\bar C}u_A^B\bar u^{\bar C}_A\rangle}}\\
\frac{u_A^{T^3}}{\sqrt{ \langle g_{B\bar C}u_A^B\bar u^{\bar C}_A\rangle}}\\
\frac{u_A^{Z}}{\sqrt{ \langle g_{B\bar C}u_A^B\bar u^{\bar C}_A\rangle}}\\
\end{matrix}
\right)\ ,\quad A=2,3,4,5\ .
\end{equation}
See eqns. \eqref{eq:matrix_solada} and \eqref{eq:matrix_sol}.
The orthogonality conditions developed in Appendix B, and presented in eqns. \eqref{orthog1} and  \eqref{orthog2}, constrain the sets of parameters which need to be fixed by the additional stabilization mechanism. 

One of these constraints is already embedded in the form of the column and row vectors; namely that they have normalize to unity.
\begin{equation}
\label{TS_orto1}
(\underline{g\cdot u_A})\cdot \overline{\underline{u_A}}=1\ .
\end{equation}
These equations show that the 20 parameters we have introduced in \eqref{TS_row_vect} and \eqref{TS_column_vect}, can be expressed in terms of only 16 independent parameters. When we further demand that these vectors are each orthogonal to the first row of the $[U_{A}^{B}]$ matrix defined in \eqref{first_row}, that is
\begin{equation}
\label{TS_orto2}
(\underline{g\cdot k})\cdot \overline{\underline{u_A}}=0\ ,
\end{equation}
we reduce the number of free parameters further, to only 12. We have defined the vector 
{\small
\begin{equation}
\begin{split}
{\underline{g\cdot k}}=\left(\frac{\langle g_{S\bar A}\bar k^{\bar B}\rangle}{\sqrt{\langle g_{B\bar C}k^B\bar k^{\bar C}\rangle}}\ ,
\frac{\langle g_{T^1\bar B}\bar k^{\bar B}\rangle}{\sqrt{\langle g_{B\bar C}k^B\bar k^{\bar C}\rangle}}\ ,
\frac{\langle g_{T^2\bar B}\bar k^{\bar B}\rangle}{\sqrt{\langle g_{B\bar C}k^B\bar k^{\bar C}\rangle}}\ ,
\frac{\langle g_{T^3\bar B}\bar k^{\bar B}\rangle}{\sqrt{\langle g_{B\bar C}k^B\bar k^{\bar C}\rangle}}\ ,
\frac{\langle g_{Z\bar B}\bar k^{\bar B}\rangle}{\sqrt{\langle g_{B\bar C}k^B\bar k^{\bar C}\rangle}}\right)\ ,
\end{split}
\end{equation} 
}
\noindent 
which projects the scalar perturbations $\delta S$, $\delta T^i$, $\delta Z$ onto the direction of $\xi^1$. 
Finally, demanding that $\underline{u_2}$, $\underline{u_3}$, $\underline{u_4}$ and $\underline{u_5}$ are also orthogonal to each other, such that
\begin{equation}
\label{TS_orto3}
(\underline{g\cdot u_A})\cdot \overline{\underline{u_B}}=0\ ,\quad\text{if}\quad  A\neq B\ ,
\end{equation}
we are left with only six free parameters. This result agrees with the formula derived in \eqref{N_degrees}, for $N=5$.
The rotation matrices $[U_{A}^{B}]$ and $[U_{A}^{B}]^{-1}$ can be expressed in terms of these six parameters after solving the orthogonality constraints \eqref{TS_orto1}, \eqref{TS_orto2} and \eqref{TS_orto3}. Adding a moduli stabilization mechanism could eliminate these degrees of freedom from the system, leaving the mixing matrices $[U_{A}^{B}]$ and $[U_{A}^{B}]^{-1}$ completely determined.

Having computed all of the mass eigenstates corresponding to D-term stabilization with FI=0, we are ready to express the Lagrangian of the associated low-energy theory. We find that
\begin{equation}
\begin{split}
\mathcal{L}&\supset
-\partial_\mu \phi \partial^\mu \phi - m_\phi^2\phi^2+i\bar \Psi \slashed\partial \Psi+M_\Psi\bar \Psi \Psi\ 
 -\frac{1}{4g_2^2}F_{\mu\nu}F^{\mu \nu}-\frac{1}{2\langle g_2^2\rangle}m_A^2A_\mu A^{\mu} \\
 &-\sum_{A=2}^{5}\partial_\mu \xi_A\partial^\mu \bar \xi^A-\sum_{A=2}^5 i\psi_{A\xi}\slashed{\partial}\psi^{A\dag}_\xi- e^{\kappa_4^2K_T/3}\mathcal{G}_{L\bar M}D_\mu C^L D^\mu \bar C^{\bar M}
 -ie^{\kappa_4^2K_T/3}\mathcal{G}_{L\bar M}\psi^L \slashed{\mathcal{D}} \psi^{\bar M\dag}+\dots\ .
 \end{split}
 \label{ny1}
\end{equation}  
Note that we have included the kinetic terms for the matter fields $( C^{L}, \psi^{L} )$, which, for the reasons discussed above, have thus far been ignored in the calculations in this Section.
Furthermore, as done throughout this paper, we have omitted any interaction terms--preferring to focus on kinetic and mass terms only. As discussed previously, the masses $m_\phi$, $m_\Psi$ and $m_A$ are all equal with their square given in \eqref{late4}. The vector boson is defined in the "unitary" gauge
\begin{equation}
A_\mu^\prime = A_\mu-\frac{\partial_\mu \eta}{\sqrt{\langle g_{B\bar C}k^B\bar k^{\bar C}\rangle}}\ .
\end{equation}
However, for simplicity of notation, we have dropped the $'$ in \eqref{ny1}.
To conclude, we have found that the effective theory of this hidden sector contains a massive vector supermultiplet, formed by a real scalar, a vector boson, and a Dirac fermion, as well as four massless chiral multiplets $(\xi^A, \psi_\xi^A)$, for $A=2,3,4,5$. The $C_L$ fields are massless matter fields on the hidden sector which do not mix with the dilaton and moduli fields. They potentially play, however, a significantly different role in vacua for which the Fayet-Iliopoulos term is non-vanishing. We briefly discuss this in the following Section.

\section{A Specific Hidden Sector with FI $\neq$ 0}

In the previous Section, we studied the hidden sector of heterotic vacua compactified on a CY threefold with $h^{1,1}=3$; that is, with moduli chiral superfields $\tilde{S}$, $\tilde{T}^{i}, i=1,2,3$ and $\tilde Z$. The low energy matter spectrum consisted of chiral superfields $\tilde{C}^{L}, L=1,\dots, \cal{N}$. Importantly, however, we explicitly chose the scalar component VEVs $\langle S \rangle$, $\langle T^{i} \rangle$ and $\langle Z \rangle$ to be such that the Fayet-Iliopoulos term vanished; that is, FI=0. In that case, $\langle \mathcal{P} \rangle =0$--and, hence, the vacuum is $N=1$ supersymmetric--if and only if the VEVs of the matter scalar fields all satisfy $\langle C^{L} \rangle =0, L=1,\dots, \cal{N}$. 

In this Section, we again consider the hidden sector of heterotic vacua compactified on a CY threefold with $h^{1,1}=3$, with moduli chiral superfields $\tilde{S}$, $\tilde{T}^{i}$, $i=1,2,3$, $\tilde Z$ and matter spectrum $\tilde{C}^{L}, L=1,\dots, \cal{N}$. However, we now chose the scalar component VEVs $\langle S \rangle$, $\langle T^{i} \rangle$ and $\langle Z\rangle $ to be such that the Fayet-Iliopoulos term is {\it non-zero}; that is, FI$\neq$0. In this case, 
$\langle \mathcal{P} \rangle =0$--and, hence, the vacuum is $N=1$ supersymmetric--if and only if the FI term is cancelled by at least one of the matter scalar fields having a non-vanishing VEV; that is, generically $\langle C^{L} \rangle \neq0, L=1,\dots, \cal{N}$. 
More specifically, let us return to the theoretical analysis outlined in Section 2. It follows from \eqref{rain5} that, in the present case,
\begin{equation}
FI= -\frac{a\epsilon_S\epsilon_R^2}{\kappa^2_4V^{2/3}\hat R}\left( \mu(L) +\frac{\pi\epsilon_S \hat R}{V^{1/3}}
\left(\beta_i^{(2)} +  W_iz^2  \right)  l^i  \right) \neq 0 \ .
\label{book1}
\end{equation}
Then we learn from \eqref{rain3} that $\langle\mathcal{P}\rangle =0\Rightarrow$
\begin{equation}
-\frac{a\epsilon_S\epsilon_R^2}{2\kappa^2_4V^{2/3}\hat R}\left( \mu(L) +\frac{\pi \epsilon_S \hat R}{V^{1/3}}
\left(\beta_i^{(2)} +  W_iz^2  \right)  l^i  \right)+G_{L\bar M}\langle C^L\rangle  \langle \bar C^{\bar M}\rangle =0\ .
\label{book2}
\end{equation}

At this point, to further simplify our equations, we make the following mild  assumptions. It was shown in~\cite{Anderson:2010ty,Ashmore:2021xdm} that bundle deformations can be used to shift the vacuum into a region of moduli space in which the first order terms in the linear expansion in the string coupling are much smaller than the terms which appear at tree-level. Assuming that we turn on the VEVs of the $C^L$ fields in such a region, we will find that
the Killing vector $k_S$, which corresponds to the dilaton field and appears only after we include the one-loop Green-Schwarz anomaly cancellation term, is negligible compared to the Killing vectors of the K\"ahler moduli and  the $C^L$ matter fields, that is, $k_{T}^{i}$ and $k_{C}^{L}$ respectively. Therefore, we will henceforth take $k_S=0$. Furthermore, we will also neglect five-brane effects which, similarly, only appear at higher orders in $\epsilon_S$ in the effective theory. 
The D-flatness condition \eqref{book2} then simplifies to
\begin{equation}
\begin{split}
-\frac{a\epsilon_S\epsilon_R^2}{2\kappa^2_4V^{2/3}\hat R}\mu(L)+G_{L\bar M}\langle C^L\rangle \langle\bar  C^{\bar M}\rangle\ =0\ .
\end{split}
\label{book3}
\end{equation}
Recall that $\mu(L)=d_{ijk}l^ia^ja^k$ is the expression for the tree-level slope of the line bundle $L$ with an anomalous $U(1)$ structure group. 
Let us now consider the F-flatness condition. Since, in this Section $FI \neq 0$, the superpotential term, which contains the matter fields $C^L$, now becomes relevant. 
Turning on VEVs for the $C^L$ fields deforms the hidden sector line bundle, breaking the anomalous $U(1)$ symmetry. In this case, the F-flatness condition does play an important role, as it restricts the space of the possible line bundle deformations, or the so called ``branch structure''; see ~\cite{Anderson:2010ty,Ashmore:2021xdm}. 

Within this context, one can then repeat the analysis of Section 3 using the formalism presented in Section 2 and Appendix B. Since the VEVs of the matter scalars fields are now non-vanishing, the associated chiral superfields $\tilde{C}^{L}$ now mix with $\tilde{S}$ and $\tilde{T}^{i}$ as shown in Lagrangian \eqref{rain1}. Here, we will not explicitly carry out the field redefinitions required to normalize the kinetic terms and to diagonalize the mass matrix. Instead, we will simply state the results. We find that the complex scalar field $\xi^{1}$ associated with the first row of the $[U_{A}^{B}]$ matrix is again given by
\begin{equation}
\xi^{1}=\eta+i\phi\ ,
\label{book4}
\end{equation}
where, using the expressions for $k_{T}^{i}$ and $k_{C}^{L} $ given in \eqref{rain2} and the scalar field metrics given in Appendix C, we find from \eqref{black6} that
\begin{equation}
\label{book5}
 \eta=\frac{4a\epsilon_S\epsilon_R^2 g^{T}_{i\bar j} l^i\chi^{\bar{j}}
+\frac{i}{2}{G}_{L\bar M} \left( \langle C^L\rangle  {\delta \bar C^{\bar M}}- \langle \bar C^{\bar M}\rangle{\delta \bar C^{\bar M}}  \right)}{\sqrt{4a^2\epsilon_S^2\epsilon_R^4g^T_{i\bar{j}}l^il^{\bar{j}}+  {G}^\prime_{L\bar{M}}  \langle C^{L} \rangle \langle \bar C^{\bar{M}} \rangle}}+\mathcal{O}(\epsilon_S^2)\\
\end{equation}
forms the longitudinal degree of freedom of the massive vector boson $A_{2\mu}$  and from \eqref{black4} that
\begin{equation}
\phi=\frac{ 2a\epsilon_S\epsilon_R^2 g^{T}_{i\bar j} l^i\delta t^{\bar{j}}
+\frac{1}{2}{G}_{L\bar M} \left( \langle C^L\rangle  {\delta \bar C^{\bar M}}+ \langle\bar C^{\bar M}\rangle{\delta \bar C^{\bar M}}  \right)}{\sqrt{4a^2\epsilon_S^2\epsilon_R^4g^T_{i\bar{j}}l^il^{\bar{j}}+   {G^\prime}_{L\bar{M}}  \langle C^{L} \rangle \langle\bar  C^{\bar{M}} \rangle}}+\mathcal{O}(\epsilon_S^2)\ .
\label{book6}
\end{equation}
is the scalar component of the massive gauge vector supermultiplet. Note the use of the scalar metrics $G_{L\bar M}$ and $G_{L\bar M}^\prime$, defined in the equations\eqref{def_GLM} and \eqref{def_GLMp}, respectively.

Similarly, the fermion $\psi_{\xi}^{1}$ is a linear combination of the fermionic parts of the chiral moduli and matter multiplets. We find that
\begin{equation}
\psi_{\xi}^{1}=\frac{ 2ia\epsilon_S\epsilon_R^2 g_{i\bar j}^T l^{\bar{j}} \psi^i_T +
i{G}_{L\bar M}   \langle \bar C^{\bar M}\rangle \psi^L}
{\sqrt{4a^2\epsilon_S^2\epsilon_R^4g^T_{i\bar{j}}l^il^{\bar{j}}+  { G^\prime}_{L\bar{M}}  \langle C^{L} \rangle \langle \bar C^{\bar{M}} \rangle}}\ .
\label{book7}
\end{equation}

As discussed in Section 2,  $\psi^1_{\xi}$ combines with the gaugino $\lambda_{2}$ to form a massive Dirac fermion
\begin{equation}
\label{book8}
\Psi= \dbinom{\lambda_{2}^{\dagger}/\langle g_2\rangle}{\psi_{\xi}^{1}} \ .
\end{equation}
We find that the masses of the fields $\phi$, $ A_{2\mu}$, and $ \Psi $ are all identical. Hence, as discussed in Section 2, they form the components of a massive vector superfield
\begin{equation}
\label{book9}
(\phi, A_{2\mu}, \Psi )
\end{equation}
whose squared mass is given by
\begin{equation}
m_{A}^{2}=m_{\text{anom}}^{2}+m_{\text{matter}}^{2},
\label{book10}
\end{equation}
where, using \eqref{repair11}, 
\begin{equation}
m_{\text{anom}}^{2} \simeq \frac{8\pi \hat{\alpha}_{\text{GUT}}}{a\text{Re}f_{2} } \Big( a^{2}\epsilon_{S}^{2}\epsilon_{R}^{2}g^{T}_{i \bar{j}}l^{i}l^{\bar{j}}  \Big) 
\label{book11}
\end{equation}
is the ``anomalous'' mass due to the inhomogeneous transformation of the $T^{i}$ axions, and from \eqref{repair12}

\begin{equation}
\begin{split}
\label{book12}
m_{\text{matter}}^{2} =\frac{\pi \hat{\alpha}_{\text{GUT}}}{a\text{Re}  f_{2} } 2  {G^\prime}_{L\bar{M}}  \langle C^{L} \rangle \langle C^{\bar{M}} \rangle
\end{split}
\end{equation}
is the usual mass generated by the Higgs mechanism associated with non-zero VEVs of the matter scalars $C^{L}$.

We conclude that the chiral superfield with components $(\xi^{1}, \psi_{\xi}^{1})$ becomes part of the massive vector multiplet and is no longer in the light effective Lagrangian. Finally, the non-interacting part of the low energy Lagrangian for the remaining zero mass canonical superfields $(\xi^{A}, \psi_{\xi}^{A})$ , $A=2,\dots, N$ is given by
\begin{equation}
\mathcal{L} \supset \sum_{A=2}^{N} \Big(-\partial_{\mu}\xi^{A}\partial^{\mu} \bar{\xi}^{\bar{A}} -i\psi_{\xi}^{A}\slashed{\partial} \psi_{\xi}^{\bar{A}\dagger} \Big) \ .
\label{book13}
\end{equation}

\appendix

\section{Anomaly Cancellation Mechanism}

At lowest order in string coupling, the bosonic part of the string frame Lagrangian takes the following form~\cite{Horava:1996ma,Green:1987mn}
\begin{equation}
\begin{split}
\label{eq:initial_act}
S_{\text{het}}&=\frac{1}{2\kappa^2_{10}}\int_{\mathcal{M}_{10}}e^{-2\phi}\left[R+4d\phi\wedge \star\phi-\frac{1}{2}H\wedge \star H
\right]\\
&-\frac{1}{2\kappa^2_{10}}\frac{\alpha^\prime}{4}\int_{\mathcal{M}_{10}} e^{-2\phi}\text{tr}(\mathcal{F}_1\wedge \star \mathcal{F}_1)+e^{-2\phi}\text{tr}(\mathcal{F}_2\wedge \star \mathcal{F}_2)\ .
\end{split}
\end{equation}
In the above action, $\phi$ is the 10D dilaton, $\mathcal{F}_1=d\mathcal{A}_1-i\mathcal{A}_1\wedge \mathcal{A}_1$ is the field strength on the observable sector, $\mathcal{F}_2=d\mathcal{A}_2-i\mathcal{A}_2\wedge \mathcal{A}_2$ is the field strength on the hidden sector and $H$ is the heterotic three-form
strength 
\begin{equation}
H=dB^{(2)}-\frac{\alpha^\prime}{4}(\omega_{YM}-\omega_L)\ .
\end{equation}
 $B^{(2)}$ is the Kalb-Ramond two-form. $\omega_{YM}$ and $\omega_L$ are the Chern-Simons three-forms defined in terms of 
the gauge potentials $\mathcal{A}_1$, $\mathcal{A}_2$ and the spin connection $\Omega$ by
\begin{equation}
\begin{split}
 d \omega_{YM}&=\text{tr}\mathcal{F}_1^2+\text{tr}\mathcal{F}_2^2\ ,\\
d \omega_{L}&=\text{tr}R^2\ .
\end{split}
\end{equation}
From the kinetic term of the three-form $H$ in \eqref{eq:initial_act}, that is
\begin{equation}
S_{\text{kin}}=-\frac{1}{4\kappa_{10}^2}\int_{\mathcal{M}_{10}}e^{-2\phi_{10}}H\wedge \star H\ ,
\end{equation}
we obtain
\begin{equation}
\begin{split}
\label{eq:kinetic_full}
S_{\text{kin}}&=-\frac{1}{4\kappa_{10}^2}\int_{\mathcal{M}_{10}}e^{-2\phi_{10}}dB^{(2)}\wedge \star dB^{(2)}
\\&+\frac{\alpha^\prime}{8\kappa_{10}^2}\int_{\mathcal{M}_{10}}(\text{tr}\mathcal{F}_1^2+\text{tr}\mathcal{F}_2^2-\text{tr}R^2)\wedge B^{(6)}+\mathcal{O}({\alpha^\prime}^2)\ .
\end{split}
\end{equation}
To obtain the result above we have used integration by parts and the duality
\begin{equation}
dB^{(6)}=e^{-2\phi}\star_{10}dB^{(2)}\ ,
\end{equation}
which relates the Kalb-Ramond two form $B^{(2)}$ to a six-form $B^{(6)}$.

Let us now assume that on the 6D compactification manifold of each sector we turn on a bundle with structure group $G^{(\alpha)}$, $\alpha=1,\>2$, such that the unbroken gauge group in 4D is $H^{(\alpha)}$. We can then write the ten-dimensional gauge field strengths $\mathcal{F}^\alpha$,  as $\mathcal{F}^\alpha\equiv F^\alpha+\bar F^\alpha$, where
$F^\alpha$ is the external four dimensional part taking values in the low energy gauge group $H^{(\alpha)}$ and 
$\bar F^\alpha$ denotes the internal six-dimensional part, which takes values in the structure group $G^{(\alpha)}$ of the
bundle. We do a similar decomposition for the gauge potentials $\mathcal{A}^\alpha=A^\alpha+\bar A^\alpha$. In the present work, we turn on a non-abelian bundle $G^{(1)}$ on the observable sector, leading to a low energy group $H^{(1)}$ in four dimensions. However, on the hidden sector, we turn on a $G^{(2)}$ bundle which contains a $U(1)$ sub-bundle. Note that there is a $U(1)$ factor in both the internal structure group of the hidden sector bundle, and in its effective theory. This type of $U(1)$ factor leads to an anomaly in the effective theory which is canceled via the four dimensional equivalent of the well-known ten dimensional Green-Schwarz mechanism~\cite{Green:1984sg}. Such a $U(1)$ is called an ``anomalous'' $U(1)$~\cite{Dine:1987xk,Dine:1987gj,Anastasopoulos:2006cz,Green:1987mn}. We denote by $f$ the $U(1)$ field strength associated to the $U(1)$ gauge connection from the low energy theory on the hidden sector, and by $\bar f$ the internal $U(1)$ field strength.

Now note that the second term in \eqref{eq:kinetic_full} can be expressed as
\begin{equation}
\begin{split}
&\frac{\alpha^\prime}{8\kappa_{10}^2}\int_{\mathcal{M}_{10}}(\text{tr}\mathcal{F}_1^2+\text{tr}\mathcal{F}_2^2-\text{tr}R^2)\wedge B^{(6)}\\
&=\frac{\alpha^\prime}{8\kappa_{10}^2}\int_{\mathcal{M}_{10}}\left(\text{tr}F_1^2+\text{tr}\bar F_1^2+2\text{tr}(F_1\bar F_1)+\text{tr}F_2^2+\text{tr}\bar F_2^2+2\text{tr}(F_2\bar F_2)-\text{tr}R^2\right)\wedge B^{(6)}\ .
\end{split}
\end{equation}
On the observable sector $\text{tr}(F_1\bar F_1)$ vanishes, because the non-abelian group $G^{(1)}$ and its commutant do not share any generators of $E_8$. On the other hand, on the hidden sector, because of the presence of the ``anomalous'' $U(1)$, we get
\begin{equation}
\text{tr}(F_2\bar F_2)=(\text{tr}\>Q^2)\>f\wedge \bar f \neq 0\equiv 4 a f\wedge \bar f\ ,
\end{equation} 
where $Q$ is the $E_8$ generator that the $U(1)$ bundle and the low energy $U(1)$ connection share.  Keeping this cross-term only, we then find
\begin{equation}
\begin{split}
\label{eq:terms_kin_reduce}
S_{\text{kin}}&=-\frac{1}{4\kappa_{10}^2}\int_{\mathcal{M}_{10}}e^{-2\phi_{10}}H\wedge \star H \\
&\supset-\frac{1}{4\kappa_{10}^2}\int_{\mathcal{M}_{10}}e^{-2\phi_{10}}dB^{(2)}\wedge \star dB^{(2)}+\frac{\alpha^\prime}{8\kappa_{10}^2}\int_{\mathcal{M}_{10}}8a\left(f\wedge \bar f\right)\wedge B^{(6)}\ .
\end{split}
\end{equation}

In this paper we will consider the compactification manifold is a Calabi-Yau (CY) threefold $X$. For the purpose of reducing from 10D to 4D, by integrating over the Calabi-Yau $X$, it is convenient to use a basis of K\"ahler $(1,1)$-forms $\omega_i$, $i=1,\dots, h^{1,1}$ and their 
Hodge duals $\hat\omega^i$ such that
\begin{equation}
\int_X\omega_i\wedge \hat \omega^j=\delta_i^j\ .
\end{equation}
Following \cite{Lukas:1999nh} (see also\cite{Brandle:2003uya,Blumenhagen:2006ux}) we reduce the first term in the sum shown in \eqref{eq:terms_kin_reduce} from 10D to 4D leads to the kinetic terms of the dilaton axion $\sigma$ and of the K\"ahler axions $\chi^i$, given by
\begin{equation}
\label{eqs:kinetic_terms}
S_{\text{kin}}\supset -\int_{M_4}\left(g_{S\bar S}\>d\sigma\wedge\star_4d\sigma+4g^T_{i\bar  j}d\chi^i\wedge\star_4 d\chi^{\bar j}\right)\ ,
\end{equation}
where 
\begin{equation}
\begin{split}
&\kappa^2_4g_{S\bar S}=\frac{1}{4V^2}\ ,\\
&\kappa_4^2g_{i\bar j}^T=-\frac{d_{ijk}t^k}{\tfrac{2}{3}d_{ijk}t^it^jt^k}+\frac{d_{ijk}t^kt^ld_{jmn}t^mt^n}{\left(\tfrac{2}{3}\right)^2(d_{ijk}t^it^jt^k)^2}\ .
\end{split}
\end{equation}
are the dilaton and K\"ahler moduli metrics, respectively.

Furthermore, dimensionally reducing the second term in the sum in \eqref{eq:terms_kin_reduce} leads to a coupling between the K\"ahler axions and the $U(1)$ gauge field $A^\mu$, given by
\begin{equation}
\begin{split}
\label{eq:1:10}
S_{\text{kin}}\subset \int_{M_4}{8a}\epsilon_S\epsilon_R^2\>g^T_{i\bar j}\chi^{\bar j}c^i_1(L)\wedge d\star_4A\ .
\end{split}
\end{equation}
Note that this coupling cannot exist unless the hidden sector contains an anomalous $U(1)$.
In the following, we will show 
that we can also find a coupling between the dilaton axion $\sigma$ and the $U(1)$ vector field in 4D, which, however, has a different origin inside the 10D theory.

It is well known that the heterotic 10D theory exhibits gravitational, gauge and mixed gauge-gravitational anomalies resulting from anomalous hexagon diagrams at one-loop in string perturbation theory. Non-factorisable anomalies vanish by themselves and the factorizable ones are canceled by adding a one-loop counter term. The Green-Schwarz anomaly canceling one-loop counter term is given by
\begin{equation}
S_{GS}=\frac{1}{48(2\pi)^5\alpha^\prime}\int_{\mathcal M_{10}}B^{(2)}\wedge X_8\ ,
\end{equation}
where the eight-form $X_8$ defined as
\begin{equation}
\begin{split}
X_8=&\frac{1}{4}(\text{tr}\mathcal{F}^2_1)^2+\frac{1}{4}(\text{tr}\mathcal{F}^2_2)^2-\frac{1}{4}(\text{tr}\mathcal{F}^2_1)(\text{tr}\mathcal{F}^2_2)\\
&-\frac{1}{8}(\text{tr}\mathcal{F}^2_1+\text{tr}\mathcal{F}^2_2)(\text{tr}R^2)+\frac{1}{8}\text{tr}R^4+\frac{1}{32}(\text{tr}R^2)^2\ .
\end{split}
\end{equation}
 As shown in ~\cite{Weigand:2006yj}, splitting further each field strength into its internal and low-energy parts, one can find that the Green-Schwarz term contains a term of the type
\begin{equation}
\label{eq:GS_term_axion}
S_{GS}\supset\sum_{\alpha=1}^2\frac{1}{4(2\pi)^3\alpha^\prime}\int_{\mathcal{M}_{10}}B^{(2)}\wedge{\text{tr}}(F_\alpha\bar F_\alpha)
\left[ \frac{1}{4(2\pi)^2}\left( \text{tr}\bar F_\alpha^2-\frac{1}{2}\text{tr}\bar R^2 \right)\right]\ ,
\end{equation}
where the sum runs over the observable and the hidden sectors. We have already explained that for our observable sector, $\text{tr}(F_1\bar F_1)$ vanishes, while on the hidden sector, which contains an anomalous $U(1)$ symmetry, $\text{tr}(F_2\bar F_2)=4a(f\wedge \bar f)$.
Then, \eqref{eq:GS_term_axion} becomes
\begin{equation}
\label{eq:GS_term_axion2}
S_{GS}\supset\frac{1}{4(2\pi)^3\alpha^\prime}\int_{\mathcal{M}_{10}}4aB^{(2)}\wedge f\wedge \bar f
\left[ \frac{1}{4(2\pi)^2}\left( \text{tr}\bar F_2^2-\frac{1}{2}\text{tr}\bar R^2 \right)\right]\ ,
\end{equation}
Reducing \eqref{eq:GS_term_axion} from 10D to 4D by integrating over the Calabi-Yau $X$, we find a term which couples the dilaton axion to the $U(1)$ gauge field in 4D:
\begin{equation}
\begin{split}
\label{eq:1.16}
S_{GS}\supset\int_{M_4}2g_{S\bar S}\>\pi a\epsilon_S^2\epsilon_R^2\beta_i \sigma c_1^i(L) \wedge d \star_4 A\ ,\quad i=1,\dots,h^{1,1}\ ,
\end{split}
\end{equation}
where $\beta^i$ are the integer charges on the hidden sector under consideration
\begin{equation}
\beta_i=-\frac{1}{v^{1/3}}\frac{1}{4(2\pi)^2}\int_X\left( \text{tr}\bar F_2^2-\frac{1}{2}\text{tr}\bar R^2 \right)\wedge \omega_i\ .
\end{equation}
The coupling between the dilaton axion and the $U(1)$ gauge field appears at one-loop in string theory, when the term responsible for the cancellation of the hexagonal diagrams in the $E_8\times E_8$ heterotic theory is included.

Combining \eqref{eqs:kinetic_terms} and \eqref{eq:1.16}, we find the following action for the dilaton axion 
\begin{equation}
\begin{split}
\label{eq:action_dilaton}
S_\sigma&=-\int_{M_4}g_{S\bar S}\>d\sigma\wedge\star_4d\sigma
+\int_{M_4}g_{S\bar S}\>2\pi a\epsilon_S^2\epsilon_R^2\beta_i \sigma c_1^i(L)\wedge d\star_4A\\
&=\int_{M_4}d^4x\sqrt{-g}\>g_{S\bar S}\>\left[-\partial_\mu\sigma\partial^\mu\sigma
+2\pi a\epsilon_S^2\epsilon_R^2 \beta_i l^i \sigma(\partial_\mu  A^\mu) \right]\ ,
\end{split}
\end{equation}
whereas for the K\"ahler axions, after we combine \eqref{eqs:kinetic_terms} and \eqref{eq:1:10}, we get
\begin{equation}
\begin{split}
\label{eq:action_axion}
S_{\chi}&=-\int_{M_4}4g^T_{i \bar j}\>d\chi^i\wedge\star_4 d\chi^{\bar j}+\int_{M_4}{8a}\epsilon_S\epsilon_R^2\>g^T_{i\bar j}\>\chi^ic_1^{\bar j}(L)\wedge d \star_4 A\\
&=\int_{M_4}d^4x\sqrt{-g}\>4g^T_{i\bar  j}\left[-\partial_\mu \chi^i\partial^\mu \chi^{\bar j}+\> 2a \epsilon_S\epsilon_R^2l^i\chi^{\bar j} (\partial_\mu A^\mu) \right]\ .
\end{split}
\end{equation}
The couplings of the axions $\sigma$, $\chi^i$ to the anomalous $U(1)$ gauge field in the effective theory induces transformation laws for these axions under the $U(1)$ symmetry.

Equations \eqref{eq:action_dilaton} and \eqref{eq:action_axion} have the generic form
\begin{equation}
\begin{split}
\label{eq:general_axion_action}
S_{\rho}&=\int_{M_4}d^4x\sqrt{-g}\left[- g_{ab}\partial_\mu \rho^a\partial^\mu\rho^b-2g_{ab}Q^a \rho^b (\partial_\mu A^\mu)\right]\\ 
&=\int_{M_4}d^4x\sqrt{-g}\left[- g_{ab}\partial_\mu \rho^a\partial^\mu\rho^b+2g_{ab}Q^a (\partial_\mu \rho^b)  A^\mu\right]\ .
\end{split}
\end{equation}
where the index $a$ runs over the dilaton and the $h^{1,1}$  K\"ahler moduli, that is $a=S, T^1, \dots, T^{h^{1,1}}$. We have used integration by parts to obtain the second line. The coefficients $Q^a$ can be read off from \eqref{eq:action_dilaton} and \eqref{eq:action_axion}. They depend on the coupling parameters $\epsilon_S$ and $\epsilon_R$ which, generically, depend on the values of the moduli. However, when evaluating these coefficients in our context, the values of the moduli are assumed fixed at a specified supersymmetric vacuum. Therefore, within our context, the coupling coefficients $\epsilon_S$, $\epsilon_R$ and, hence, $Q^a$ are non-dynamical fields and can be considered to be constants. Similarly, we assumed that the moduli space metric $g_{a\bar b}$ is fixed at the vacuum; that is, $g_{a\bar b}= \langle g_{a\bar b}\rangle$. Perturbations of the metric around the vacuum are neglected in this analysis. 
It can be also be shown that higher order contributions $(\mathcal{O}(\alpha^\prime))$ lead to terms of the type $-Q_a^2A^\mu A_\mu$ for  $a=S, T^1, \dots, T^{h^{1,1}}$; see discussions in ~\cite{Lukas:1999nh,Ibanez:2012zz}. These can be added to the generic action in \eqref{eq:general_axion_action} to obtain
\begin{equation}
\begin{split}
\label{eq:general_axion_action_full}
S_{\rho}&=\int_{M_4}d^4x\sqrt{-g}\left[- (\partial^\mu\rho_a)^2+2Q^a (\partial_\mu \rho_a)  A^\mu
-Q_a^2A^\mu A_\mu\right]\\
&=-\int_{M_4}d^4x\sqrt{-g}\left[\partial_\mu\rho_a-Q_a   A_\mu\right]^2\ .
\end{split}
\end{equation}
If the anomalous $U(1)$ symmetry of the low energy theory is gauged and the gauge vector field $A^\mu$ transforms as
\begin{equation}
A^\mu \rightarrow A^\mu+\partial^\mu \theta\ ,
\end{equation}
 then it can be shown that the action $S_{\rho}$ is gauge invariant only if the scalar fields $\rho^a$ transform as
\begin{equation}
\rho^a\rightarrow \rho^a+Q^a \theta\ .
\end{equation}
In the transformations above, $\theta=\theta(x)$ is an arbitrary gauge parameter. 
Comparing the generic case shown in equation \eqref{eq:general_axion_action}, to the actions 
\eqref{eq:action_dilaton} and \eqref{eq:action_axion}, involving the dilaton axion the and K\"ahler axions respectively, we learn that the axions in our low energy theory have the following transformations under the anomalous $U(1)$:
\begin{equation}
\begin{split}
\label{eq:axions_plm}
&\sigma \rightarrow \sigma+Q^0\theta\ ,\quad Q^0=-\pi a\epsilon_S^2\epsilon_R^2 \beta_i l^i\ ,\\
&\chi^i\rightarrow \chi^i+Q^i\theta\ ,\quad Q^i=-\pi a \epsilon_S\epsilon_R^2l^i\ , \quad i=1,\dots,h^{1,1}\ .
\end{split}
\end{equation}

These set of axions represent the imaginary components of the dilaton moduli $S$ and the K\"ahler moduli $T$. As shown in Section 2, the expressions for these fields in the presence of five-branes are
\begin{equation}
\begin{split}
\label{eq:def_scalar}
& S=V+\pi\epsilon_SW_it^i\left(\tfrac{1}{2}+\lambda\right)^2
+i\left[\sigma+2 \pi\epsilon_S W_i\chi^iz^2\right] \ ,\\
&T^i=t^i+2i\chi^i\ ,\quad i=1,\dots,h^{1,1}\ ,
\end{split}
\end{equation}
where $W_i$ and $\lambda_i$ specify properties of the internal five-brane.
Hence, the $S$ and $T^i$ moduli have the following transformations under $U(1)$:
\begin{equation}
\begin{split}
\label{eq:def_scalar_transform}
& \delta_\theta S=iQ^0\theta+2i\pi\epsilon_S Q^iW_iz^2  \theta \equiv k_S\theta,\\
&\delta_\theta T^i=2iQ^i \theta\equiv k_T^i\theta\ ,\quad i=1,\dots,h^{1,1}\ ,
\end{split}
\end{equation}
where we have defined
\begin{equation}
\begin{split}
\label{eq:def_scalar_killing}
 k_S&=iQ^0+2i\epsilon_S Q^iW_iz^2  \\
&=-2i\pi a\epsilon_S^2\epsilon_R^2\left( \tfrac{1}{2}\beta^{N+1}_i l^i + W_il^iz^2\right)\\
k_T^i&=2iQ^i=-2i a\epsilon_S\epsilon_R^2l^i ,\quad i=1,\dots,h^{1,1}\ .
\end{split}
\end{equation}
Note that the five-brane modulus,
\begin{equation}
Z=W_it^iz+2iW_i(-\eta^i\nu+\chi^iz)\ ,
\end{equation}
contains the K\"ahler axions in the definition of its imaginary component. Therefore, the five-brane modulus transforms inhomogenously under under $U(1)$ as well, 
\begin{equation}
\delta_\theta Z=2iW_iz Q^i\theta=k_Z\theta\ ,
\end{equation}
where we defined
\begin{equation}
k_Z=2iW_iz Q^i=-2ia\epsilon_S\epsilon_R^2W_il^iz\ .
\end{equation}
Hence, we have obtained the inhomogenous gauge transformations of the $S$, $T^i$ and $Z$ moduli fields, and defined them in terms of the Killing vectors $k_S$, $k_T^i$ and $k_Z$.

\section{Invariant Moduli/Matter Lagrangians under Anomalous $U(1)$ Transformations --General Formalism}

The effective theories we are interested in discussing in this paper  are best analyzed using the formalism of the non-linear $N=1$ SUSY $\sigma-$model; see, for example, the analysis shown in~\cite{Freedman:2012zz}.
The role of this Appendix is to outline the generic characteristics of this formalism, for a set of chiral superfields which transform under an anamolous $U(1)$ symmetry. We will consider a set of $N$ complex chiral superfields $Z^A$, with scalar, fermionic and auxiliary field components $(z^A, \psi^A, F^A)$, coupled to the same $U(1)$ gauge supermultiplet. We will formulate the F-term and D-term stabilization conditions and, hence, show that, in general, the D-term stabilization mechanism can lead to non-zero VEVs for the scalar components and, consequently, to the formation of a massive $U(1)$ vector superfield.

We begin our general analysis by assuming a K\"ahler potential $K(z,\bar z)$ and a K\"ahler metric
\begin{equation}
g_{A\bar B}=\partial_A\partial_{\bar B}K(z,\bar z)\ ,
\end{equation}
which has a Lie group of symmetries generated by holomorphic Killing vectors. Under these symmetries, the components of the chiral multiplets have the following infinitesimal transformations~\cite{Wess:1992cp,Freedman:2012zz}:
\begin{equation}
\begin{split}
&\delta_\theta z^A=k^A(z)\theta\ ,\\
&\delta_\theta \psi^A= \frac{\partial k(z)^A}{\partial z^B}\psi^B\theta\,\\
&\delta_\theta F^A=\frac{\partial k^A(z)}{\partial z^B}F^B\theta-
\frac{1}{2}\frac{\partial^2k(z)^A}{\partial z^B \partial z^C}\psi^{B\dag}\psi^C\theta\ .
\end{split}
\end{equation}
The vectors $k^A(z),\>\bar k^{\bar A}(\bar z)$ are related to real scalar moment map $\mathcal{P}(z,\bar z)$, such as
\begin{equation}
k^A=-ig^{A\bar B}\partial_{\bar B}\mathcal{P}(z, \bar z)\ , \quad \bar k^{\bar A}=-ig^{A\bar B}\partial_{A}\mathcal{P}(z, \bar z)\ .
\end{equation}
Inverting the expressions above we get
\begin{equation}
\label{eq:Killing_vectors}
\mathcal{P}(Z,\bar Z)=ik^A\partial_AK(z,\bar z)=-i\bar k^{\bar A}\partial_{\bar A}K(z,\bar z)\ .
\end{equation}
In the following we will show that the D-term flatness condition is naturally expressed in terms of this moment map.

As we promote this Killing symmetry to a gauge symmetry, with parameter $\theta\rightarrow \theta(x)$, 
we introduce a $U(1)$ gauge supermultiplet $(A_\mu,\lambda,D)$ with gauge coupling constant $g$. The next step is to define covariant derivatives for the fields in the $N$ chiral multiplets $Z^A=(z^A, \psi^A, F^A)$,
\begin{equation}
\begin{split}
&\mathcal{D}_\mu z^A=\partial_\mu z^A-A_\mu k^A\ ,\\
&\mathcal{D}_\mu \bar z^{\bar A}=\partial_\mu \bar z^{\bar A}-A_\mu \bar k^{\bar A}\ ,\\
&\mathcal{D}_\mu \psi^A=\partial_\mu \psi^A-A_\mu \frac{\partial k^A}{\partial z^B}\psi^B\ .
\end{split}
\end{equation}
These covariant derivatives are used to build a Lagrangian with $N$ chiral superfields $Z^A$, which is invariant under the gauged anomalous $U(1)$ symmetry. The result is
\begin{equation}
\label{eq:initial_lagrangian}
\begin{split}
\mathcal{L}&\supset -g_{A\bar B}\mathcal{D}_\mu  z^A \mathcal{D}^\mu \bar z^{\bar B}  -ig_{A\bar B}\psi^A \slashed{\mathcal{D}} \psi^{\bar B\dag}- \frac{i}{g^2}\lambda \slashed{\partial} \lambda^\dag -\frac{1}{4g^2}F_{\mu\nu}F^{\mu \nu}
+\sqrt{2}g_{A\bar B}k^A\lambda^\dag \psi^{\bar B\dag}\\&+\sqrt{2} g_{A\bar B}\bar  k^{\bar B}\lambda \psi^A
-\frac{1}{2}g^2\mathcal{P}^2-g^{A\bar B}\frac{\partial \mathcal{W}}{\partial z^A}\frac{\partial\mathcal{\bar W}}{\partial \bar z^{\bar B}}\dots\ , \quad A,B=1,\dots, N\ .\\
\end{split}
\end{equation}
The coupling $g=g(z)$ is the $U(1)$ gauge coupling. It can be expressed in terms of the holomorphic gauge kinetic function $f(z)$ as
\begin{equation}
\frac{1}{g^2(z)}=\text{Re}f(z)\ .
\end{equation} 
The scalar potential contains a D-term potential $V_D$, defined in terms of the real moment maps and an F-term potential $V_F$, defined in terms of the superpotential $\mathcal{W}(Z)$.
\begin{equation}
 V_D+V_F=\frac{1}{2}g^2\mathcal{P}^2+g^{A\bar B}\frac{\partial \mathcal{W}}{\partial z^A}\frac{\partial\mathcal{\bar W}}{\partial \bar z^{\bar B}}\ .
\end{equation}

Unbroken supersymmetry requires both D-term and F-term potentials to vanish. These requirements are also called the D-flatness and F-flatness conditions. More explicitly, in a supersymmetric vacuum
\begin{equation}
\langle \mathcal{P}\rangle =0\ , \quad \langle  \frac{\partial \mathcal{W}}{\partial z^A} \rangle=0\ , \quad A=1,\dots, N\ .
\end{equation}
In general, these conditions alone do not fix all the VEVs of the scalar fields $z^A$ of the system. They do, however, restrict the possible range that the VEVs of the scalar fields $z^A$ can obtain. 
The D-flatness condition can be written
\begin{equation}
\label{eq:susy_condition_0}
\langle \mathcal{P}\rangle= \langle\> k^A(z)\partial_A K(z,\bar z)\>\rangle =0\ .
\end{equation}
The condition above does not lead to particularly interesting effects if the effective theory contains only chiral fields which transform homogeneously under the $U(1)$ symmetry. Indeed, when this is the case, $k^A\sim z^A$, and hence, the D-term potential always vanishes when the values of the scalar fields are equal to zero. Such scalars would not aquire non-zero VEVs. On the other hand, if the theory contains fields which transform inhomogenously under the $U(1)$ symmetry, the D-flatness condition can lead to non-trivial VEVs of the scalar fields. 

Let us assume that among the fields $z^A$ we find some which do transform inhomogenously under $U(1)$. In the following we will prove that in a D-flat, non-trivial vacuum, a massive vector supermultiplet is always produced. To display this process, we expand the scalar fields around
the vacuum $z^A=\langle z^A\rangle +\delta z^A$. Note that fixing the VEVs $\langle z^A \rangle$ results in fixing an 
expectation value for the $z$-dependent Killing vectors, as well as for the $z$-dependent K\"ahler metric $g_{A\bar B}$ and the gauge coupling $g$. We use the notation
\begin{equation}
\begin{split}
&\langle k^A(z)\rangle =\left.k^A(z)\right|_{ z=\langle z \rangle}\ , \quad \langle \bar k^{\bar A}(\bar z)\rangle =\left.\bar k^{\bar A}(\bar z)\right|_{ \bar z=\langle \bar z \rangle}\ ,\\
&\langle g_{A\bar B}(z,\bar z)\rangle =\left.g_{A\bar B}(z,\bar z)\right|_{ z,\bar z=\langle z\rangle, \langle \bar z\rangle}\ ,\quad \langle g(z)\rangle =\left.g(z)\right|_{ z=\langle z \rangle}\ .
\end{split}
\end{equation}
which means that the metric and the Killing vectors are evaluated in the vacuum delimited by the D-flatness condition.

In $N=1$ SUSY, the massive vector multiplet has one spin-$0$ component, two spin-$\frac{1}{2}$ components and one spin-$1$ component~\cite{Wess:1992cp,Shifman:2012zz}. More specifically, these components are: a real scalar, a massive vector boson and a Dirac fermion. Next, 
we will attempt to identify the origin of each of these elements.

\begin{itemize}

\item Massive real scalar $\phi$:

Expanding around the (assumed non-trivial) vacuum defined in \eqref{eq:susy_condition_0} we get:
\begin{equation}
\delta\mathcal{P}=\langle \frac{\partial \mathcal{P}}{\partial z^A} \rangle \delta z^A+\langle \frac{\partial \mathcal{P}}{\partial {\bar z}^{\bar B}}\rangle \delta z^{\bar B}\ .
\end{equation}
Now, inverting the equations from \eqref{eq:Killing_vectors} we find
\begin{equation}
\partial_{\bar B}\mathcal{P}=ik^Ag_{A\bar B}\ ,\quad \partial_{A}\mathcal{P}=-ik^{\bar B}g_{A\bar B}\ ,
\end{equation}
and therefore, we express the perturbation of the moment map as
\begin{equation}
\delta\mathcal{P}=i\langle k^{\bar B}g_{A\bar B}\rangle \delta z^A-i\langle k^Ag_{A\bar B}\rangle \delta \bar z^{\bar B}\equiv -2\sqrt{\langle g_{A\bar B}k^A\bar k^{\bar B}\rangle}\phi \ .
\end{equation}
In the expression above, we defined a scalar field $\phi $, as 
\begin{equation}
\label{eq:phi_def}
\phi=\frac{ i\langle k^Ag_{A\bar B}\rangle \tfrac{\delta \bar z^{\bar B}}{2}-i\langle k^{\bar B}g_{A\bar B}\rangle \tfrac{\delta z^A}{2}  }{\sqrt{ \langle g_{A\bar B}k^A\bar k^{\bar B}\rangle}}\\
\end{equation}
Hence, we found that as we expand around the vacuum, we obtain the following expression for the potential energy
\begin{equation}
\label{eq:potential_phi}
\mathcal{L}\supset- \frac{1}{2}\delta (g^2)\langle \mathcal{P}^2\rangle- \frac{1}{2}\langle g^2\rangle (\langle \mathcal{P}\rangle+\delta \mathcal{P})^2=-\frac{1}{2}\langle g^2\rangle \delta \mathcal{P}^2=-2\langle g^2g_{i\bar j}k^A\bar k^{\bar B}\rangle\phi^2\ ,
\end{equation}
which is a mass term for the real scalar $\phi$. 
The field $\phi$ has a canonically normalized kinetic term in the Lagrangian. To prove it, it is best to rotate the basis of scalar perturbations $\{\delta z^A\}$  into a new basis of scalars $\{ \xi^A\}$, which have canonically normalized kinetic energy. We interpret the fields $\xi^A$, $A=1,\dots, N$ to be the true mass eigenstates of the system. However, in the vacuum defined by the D-flatness condition, only one such eigenstate, which we denote by $\xi^1$, becomes massive. The rest of the scalars $\xi^A$, $A=2,\dots,N$ fields remain massless. 

The two sets of fields, $\{ \xi^A\}$ and $\{ \delta z^A\}$, $A=1,\dots, N$, are related by the rotation matrix $U$, such that
\begin{equation}
\begin{split}
&\xi^A=[U]^A_B\delta z^B\ , \quad \bar \xi^A=[U^*]^{\bar A}_{\bar B}\delta\bar z^B\ ,\\
&\delta z^A=[{U^{-1}}]^A_B\xi^B\ , \quad \delta\bar z^{\bar A}=[{U^{*-1}}]^{\bar A}_{\bar B}\bar\xi^{\bar B}\ .
\end{split}
\end{equation}
%
We demand that after the rotation, the fields $\xi^A$ have canonically normalized kinetic energy and therefore
\begin{equation}
\langle g_{A\bar B}\rangle \partial_\mu \delta z^A \partial^\mu\delta \bar z^{\bar B}=\langle g_{A\bar B}\rangle[{U^{-1}}]^A_C\bar [{U^{-1}}]^{\bar B}_{\bar D}
\partial_\mu  \xi^C \partial^\mu \bar \xi^{\bar D}=\delta_{C\bar D}\partial_\mu  \xi^C\partial^\mu\ \bar \xi^{\bar D}\ ,
\end{equation}
which is possible only if
\begin{equation}
\label{eq:matrix_condition_x}
\langle g_{A\bar B}\rangle[{U^{-1}}]^A_C [{U^{*-1}}]^{\bar B}_{\bar D}=\delta_{C\bar D}\ .
\end{equation}
The condition shown above is not enough by itself to determine all the elements of the rotation matrix. One possible ansatz for the rotation matrix $[U^{-1}]^A_B$, which solves \eqref{eq:matrix_condition_x}, is 
\begin{equation}
\label{eq:matrix_sol}
[U^{-1}]^A_B=
\left(
\begin{matrix}
\frac{\langle k^1\rangle}{\sqrt{ \langle g_{A\bar B}k^A\bar k^{\bar B}\rangle}}&\frac{u^1_{2}}{\sqrt{\langle g_{A\bar B}u_2^A\bar u_2^{\bar B}\rangle}}&\dots&\frac{u^1_{N}}{\sqrt{\langle g_{A\bar B}u_N^A\bar u_N^{\bar B}\rangle}}\\
\frac{\langle k^2 \rangle}{\sqrt{\langle g_{A\bar B}k^A\bar k^{\bar B}\rangle}}&\frac{u^2_{2}}{\sqrt{\langle g_{A\bar B}u_2^A\bar u_2^{\bar B}\rangle}}&\dots&\frac{u^2_{N}}{\sqrt{\langle g_{A\bar B}u_N^A\bar u_N^{\bar B}\rangle}}\\
\vdots& \vdots & \ddots &\vdots\\
\frac{\langle k^N \rangle}{\sqrt{\langle g_{A\bar B}k^A\bar k^{\bar B}\rangle}}&\frac{u^N_{2}}{\sqrt{\langle g_{A\bar B}u_2^A\bar u_N^{\bar B}\rangle}}&\dots&\frac{u^N_{N}}{\sqrt{\langle g_{A\bar B}u_N^A\bar u_N^{\bar B}\rangle}}
\end{matrix}
\right)\ \equiv\left(\underline k\>,\underline{u_2},\>\dots,\>\underline{u_N} \right)\ ,
\end{equation}
The matrix can be expressed as a set of column vectors $(\underline{k}, \underline{u_n})$, $n=2,\dots,N$, as shown above (note that we considered $u_1\equiv k$).
After solving the $D$-flatness condition shown in \eqref{eq:susy_condition_0}, we can fix the vector $\underline{k}$ of the matrix only. The reason for this ansatz, as well as the physical implications that follow, will become apparent later.
The rest of the column vectors in this matrix, $\underline{u_n}$, $n=2,\dots,N$ cannot be completely determined in the absence of yet another vacuum stabilization potential in the theory. They not are not entirely unconstrained, though. They are orthogonal to the vector $\underline k$. 
To prove this, we write condition \eqref{eq:matrix_condition_x} in terms of the elements of the matrix shown above and obtain
 \begin{equation}
 \label{orthog1}
 \langle g_{A\bar B}k^A\rangle \bar u^{\bar B}_n=\delta_{1n}=0 \quad \text{for}\quad n\in[2,\dots,N]\ .
 \end{equation}
 The index $n$ runs over the subscript of the column vectors $\left(\underline{u_2},\>\dots,\>\underline{u_N} \right)$. We introduce the tensor product $\langle g_{A\bar B}k^A\rangle =({g\cdot k})_{\bar B}$, which defines the contravariant vector $(\underline{g\cdot k})$, and write
 \begin{equation}
 \label{equ:orthogo_exp}
\langle  g_{A\bar B}k^A\rangle \bar u_n^{\bar B}=({g\cdot k})_{\bar B} \bar u_n^{\bar B}=(\underline{g\cdot k})\cdot \overline{\underline{ u_n}}=\delta_{1n}\ .
 \end{equation}
Note that we have introduced the dot product between the vectors $(\underline{g\cdot k})$ and $\overline{\underline{ u_n}}$. The expression in \eqref{equ:orthogo_exp} represents a set of $N-1$ equations that express the orthogonality between each of the $N-1$ column vectors $\underline{u_n}$, $n=2,\dots,N$ and the vector $\underline{k}$.
This set of orthogonality relations is not the only one we obtain; remember that we demand that all the $N$ fields $\xi^A$ have canonically normalized kinetic terms; that is, condition \eqref{eq:matrix_condition_x} constraints the set of vectors $\underline{u_n}$, $n=2,\dots, N$ to be orthogonal to each other and have unit length:
\begin{equation}
\begin{split}
\label{orthog2}
&\langle g_{A\bar B}\rangle u_n^A\bar u_m^{\bar B}=(\underline{g\cdot u_n}) \cdot \overline{\underline{ u_m}}=\delta_{nm}=1 \quad \text{if}\quad n= m\ , \quad \text{for}\quad n,\>m\in[2,\dots,N]\ .\\
&\langle g_{A\bar B}\rangle u_n^A\bar u_m^{\bar B}=(\underline{g\cdot u_n}) \cdot \overline{\underline{ u_m}}=\delta_{nm}=0 \quad \text{if}\quad n\neq m\ , \quad \text{for}\quad n,\>m\in[2,\dots,N]\ .
 \end{split}
\end{equation}
We have defined the tensor product $g_{A\bar B}u_n^A=(\underline{g\cdot u_n})_{\bar B}$, which defines the $N-1$ contravariant vectors $(\underline{g\cdot u_n})$.

The total number of undetermined matrix entries within the set of $N-1$ column vectors $\underline{u_n}$, $n=2,\dots,N$ is $N-1\times N$. We have shown that these parameters are not completely independent, however. The orthogonality constraints \label{eq:matrix_condition} significantly reduces the number of degrees of freedom of the system. $N-1$ constraints arise because these $N-1$ column vectors have unit length ($|u_n|=1$). The requirement that these $N-1$ column vectors are orthogonal to each other, as well as to the fixed direction $\underline k$ introduces another set of $\frac{1}{2}N(N-1)$ constraints. Therefore, the real number of independent free parameters, ``hidden'' in the matrix \eqref{eq:matrix_sol} is actually equal to
\begin{equation}
\begin{split}
\label{N_degrees}
N_{\text{dof}}&=(N-1) N-(N-1)-\frac{1}{2}(N-1)\\
&=\frac{1}{2}(N-1)(N-2)\ .
\end{split}
\end{equation}
The inverse matrix $U$ can be inferred from the condition $UU^{-1}=1_N$, together with the orthogonality requirement shown in \eqref{orthog1} and \eqref{orthog2}:
\begin{equation}
\label{eq:matrix_solada}
[U]^A_B=
\left(
\begin{matrix}
\frac{\langle g_{1\bar A}\bar k^{\bar A}\rangle}{\sqrt{\langle g_{A\bar B}k^A\bar k^{\bar B}\rangle}}& \frac{\langle g_{2\bar A}\bar k^{\bar A}\rangle}{\sqrt{\langle g_{A\bar B}k^A\bar k^{\bar B}\rangle}}&\dots&\frac{\langle g_{N\bar A}\bar k^{\bar A}\rangle}{\sqrt{\langle g_{A\bar B}k^A\bar k^{\bar B}\rangle}}\\
\frac{\langle g_{1\bar A}\bar u_2^{\bar A}\rangle}{\sqrt{\langle g_{A\bar B}u_2^A\bar u_2^{\bar B}\rangle}}&\frac{\langle g_{2\bar A}\bar u_2^{\bar A}\rangle}{\sqrt{\langle g_{A\bar B}u_2^A\bar u_2^{\bar B}\rangle}}&\dots&\frac{\langle g_{N\bar i}\bar u_2^{\bar A}\rangle}{\sqrt{\langle g_{A\bar B}u_2^A\bar u_2^{\bar B}\rangle}}\\
\vdots& \vdots& \ddots&\cdots\\
\frac{\langle g_{1\bar A}\bar u_N^{\bar A}\rangle}{\sqrt{\langle g_{A\bar B}u_2^A\bar u_2^{\bar B}\rangle}}&\frac{\langle g_{2\bar A}\bar u_N^{\bar A}\rangle}{\sqrt{\langle g_{A\bar B}u_2^A\bar u_2^{\bar B}\rangle}}&\dots&
\frac{\langle g_{N\bar A}\bar u_N^{\bar A}\rangle}{\sqrt{\langle g_{A\bar B}u_2^A\bar u_2^{\bar B}\rangle}}\\
\end{matrix}
\right)\ 
\equiv\left(
\begin{matrix}
\underline{g\cdot \bar k}\\ \underline{g\cdot \bar u_2}\\ \dots \\ \underline{g\cdot \bar u_N}
\end{matrix}
\right)\ .
\end{equation}
As shown above, we can write this matrix as a set of row vectors $\underline{g\cdot k}$, $\underline{g\cdot u_n}$, for $n=2,\dots,N$, which were first encountered in the orthogonality relations \eqref{equ:orthogo_exp} and \eqref{orthog2}. As mentioned previously, the $U$ matrix combines the set of $N$ scalar linear perturbations around the vacuum, $\delta z^A$, into a set of $N$ physical states, $\xi^A$. However, the $D$-flatness condition shown in \eqref{eq:susy_condition_0} determines only the first row of this matrix and hence, we can determine exactly one physical state:
\begin{equation}
\label{eq:define_xi}
\xi^1=\frac{\langle g_{A\bar B}\bar k^{\bar B}\rangle}{\sqrt{\langle g_{A\bar B}k^A\bar k^{\bar B}\rangle}}\delta z^A\equiv (\underline{g\cdot k})\cdot \underline{\delta z}\ .
\end{equation}

It is now time to reveal why we have chosen the ansatz shown in eq. \eqref{eq:matrix_solada} and \eqref{eq:matrix_sol} for the matrix $U$ and its inverse, respectively. Let us first split the field $\xi^1$ into its real and imaginary parts.
Using the definition of $\xi^1$, one can then check that we can find the following expression
\begin{equation}
\begin{split}
\label{eq:phi_eta_def}
&\text{Re}\xi^1=\frac{\langle  g_{A\bar B}k^A \rangle  \tfrac{\delta \bar z^{\bar B}}{2}+\langle  g_{A\bar B}\bar k^{\bar B}\rangle \tfrac{\delta  z^{A}}{2}}
{\sqrt{ \langle g_{A\bar B}k^A\bar k^{\bar B}\rangle}}\ ,\\
&\text{Im}\xi^1=\frac{ i\left( \langle g_{A\bar B}k^A\rangle\tfrac{\delta \bar z^{\bar B}}{2}-\langle g_{A\bar B}\bar k^{\bar B}\rangle \tfrac{\delta z^A}{2}\right)  }{\sqrt{\langle  g_{A\bar B}k^A\bar k^{\bar B}\rangle}}\ .
\end{split}
\end{equation}
We have recovered the field $\phi=\text{Im}\xi^1$, first defined in eq. \eqref{eq:phi_def}, which represents
the direction in which we develop the quadratic potential shown in \eqref{eq:potential_phi}. This quadratic potential is responsible for stabilizing the theory to a supersymmetric vacuum. 

Note that both fields $\phi$ and $\eta$ have canonically normalized kinetic terms, that is
\begin{equation}
\mathcal{L}\supset -\partial_\mu \xi^1\partial^\mu \bar \xi^1=-\partial_\mu \phi\partial^\mu \phi-\partial_\mu \eta 
\partial^\mu \eta\ .
\end{equation}
Combining the kinetic terms of the scalars with the quadratic potential shown in \eqref{eq:potential_phi} it follows that
\begin{equation}
\mathcal{L}\supset -\partial_\mu \phi \partial^\mu \phi -\partial_\mu \eta 
\partial^\mu \eta-\partial_\mu \xi_A\partial^\mu \bar \xi^A-2\langle g^2g_{A\bar B}k^A\bar k^{\bar B}\rangle\phi^2\ ,
\quad A=2,\dots, N\ .
\end{equation}
We see that the D-term potential produced a mass term for the scalar $\phi$,
\begin{equation}
m_\phi=\sqrt{ 2\langle g^2g_{A\bar B}k^A\bar k^{\bar B}\rangle}\ ,
\end{equation}
while the other $N-1$ scalars $\xi^A$ remained massless.
After it aquires a mass, $\phi$ becomes the real scalar component of a supersymmetric massive vector multiplet. 

The real part of $\xi^1$, which we denote $\eta=\text{Re}\xi^1$ also plays an important role in the theory. In the next part we will show that $\eta$ is the Goldstone scalar which becomes the longitudinal degree of freedom of the massive vector boson $A^\mu$. All other $N-1$ states $\xi^A$, remain undetermined. We know, however, that they must be orthogonal to the direction in which we develop a potential, and hence, they remain as flat directions in the theory. The role played by the matrix $U$ and its inverse, therefore, is to project the field perturbations around the supersymmetic vacuum into these $N$ physical eigenstates $\xi^A$.

\item Massive vector boson $A^\mu$:

From the covariant derivative of the scalar field we get a mass term for the vector boson
\begin{equation}
\begin{split}
\label{eq:covariant_der_term}
\mathcal{L}\subset &  -g_{A\bar B}\mathcal{D}_\mu  z^A  \mathcal{D}^\mu \bar z^{\bar B}=-g_{A\bar B}\mathcal{D}_\mu  (\langle z^A\rangle +\delta z^A) \mathcal{D}^\mu (\langle\bar z^{\bar B}\rangle +\delta \bar z^{\bar B})\\
&=-g_{A\bar B}\partial_\mu \delta z^A \partial^\mu\delta \bar z^{\bar B}+g_{A\bar B}(\langle k^A\rangle \partial_\mu \delta \bar z^{\bar B}+\langle \bar k^{\bar B}\rangle \partial_\mu\delta  z^{A})A_\mu-\langle g_{A\bar B} k^A \bar k^{\bar B}\rangle A^\mu A_\mu\\
&=-g_{A\bar B}\partial_\mu \delta z^A \partial^\mu\delta \bar z^{\bar B}+2\sqrt{\langle g_{A\bar B}k^A\bar k^{\bar B}\rangle}\partial_\mu \eta A^\mu- \langle g_{A\bar B} k^A \bar k^{\bar B}\rangle A^\mu A_\mu \ .
\end{split}
\end{equation}

We combined the terms linear in $\partial_\mu \delta z^{A}, \partial_\mu \delta \bar z^{\bar A}$ into the field
\begin{equation}
\eta=\frac{\langle g_{A\bar B}k^A\rangle \tfrac{\delta \bar z^{\bar B}}{2}+\langle g_{A\bar B} \bar k^{\bar B}\rangle \tfrac{\delta  z^{A}}{2}}
{\sqrt{\langle g_{A\bar B}k^A\bar k^{\bar B}\rangle}}\ ,
\end{equation}
which we first encountered in eq. \eqref{eq:phi_eta_def}. We can now write \eqref{eq:covariant_der_term} in the form

\begin{equation}
\begin{split}
\mathcal{L}\subset & - g_{A\bar B}\mathcal{D}_\mu  z^A  \mathcal{D}^\mu \bar z^{\bar B}=\\
&=\dots+- \partial_\mu \eta\partial^\mu \eta+2\sqrt{\langle g_{A\bar B}k^A\bar k^{\bar B}\rangle}\partial_\mu \eta A^\mu- \langle g_{A\bar B} k^A \bar k^{\bar B}\rangle A^\mu A_\mu\\
&=\dots -\langle g_{A\bar B} k^A \bar k^{\bar B}\rangle \left(A_\mu-\frac{\partial_\mu \eta}{\sqrt{\langle g_{A\bar B}k^A\bar k^{\bar B}\rangle}}\right)\left(A^\mu-\frac{\partial^\mu \eta}{\sqrt{\langle g_{A\bar B}k^A\bar k^{\bar B}\rangle}}\right)\\
&=-\langle g_{A\bar B} k^A \bar k^{\bar B}\rangle A^\prime_\mu A^{\prime\mu}\ .
\end{split}
\end{equation}
In the above expression, we have used the ``unitary gauge'' to define the field 
\begin{equation}
A_\mu^\prime = A_\mu-\frac{\partial_\mu \eta}{\sqrt{\langle g_{A\bar B}k^A\bar k^{\bar B}\rangle}}\ .
\end{equation}
We found that after the scalar fields $z^i$ aquire VEVs, $\partial_\mu \eta/\sqrt{\langle g_{A\bar B}k^A\bar k^{\bar B}\rangle}$ becomes the longitudinal component of a now massive vector field $A_\mu^\prime$. The field $\eta$ plays the role of a Goldstone boson ``eaten'' by the vector boson. 

Including the kinetic energy term of the vector boson we have
\begin{equation}
\begin{split}
\mathcal{L}&\supset -\frac{1}{4g^2}F_{\mu\nu}F^{\mu \nu}-\langle g_{A\bar B}k^A \bar k^{\bar B}\rangle A_\mu A^{\mu}\\
&=-\frac{1}{4\langle g^2\rangle }F_{\mu\nu}F^{\mu \nu}-\langle g_{A\bar B}k^A \bar k^{\bar B}\rangle A_\mu A^{\mu}+\dots
\Rightarrow m_A^2=2\langle g^2g_{A\bar B}k^A\bar k^{\bar B}\rangle\ .
\end{split}
\end{equation}
Note that we have dropped the prime on $A_{2\mu}$ for simplicity. The dots represent higher order interaction terms obtained after expanding the gauge kinetic function around its fixed value in the vacuum.

\item Massive Dirac fermions $\Psi$:

The Lagrangian shown in \eqref{eq:initial_lagrangian} contains cross couplings between the gaugino $\lambda$ and the fermions $\psi^A$. After the fields $z^A$ obtain VEVs, these couplings mix the gaugino with the chiral fermions into a new mass eigenstate. We have
\begin{equation}
\begin{split}
\mathcal{L}\subset \sqrt{2}\langle g_{A\bar B}\rangle \langle k^A\rangle \lambda^\dag \psi^{\bar B\dag}+\sqrt{2} \langle g_{A\bar B}\rangle \langle\bar k\rangle^{\bar B}\lambda \psi^A\equiv &\sqrt{2 \langle g_{A\bar B}k^A\bar k^{\bar B}\rangle}
\left(\psi_\xi^{1\dag} \lambda^\dag+\lambda \psi^1_\xi\right)\\
=&\sqrt{2 \langle g^2 g_{A\bar B}k^A\bar k^{\bar B}\rangle}\bar \Psi \Psi\ .
\end{split}
\end{equation}
To obtain the last expression, we had to define:
\begin{equation}
\begin{split}
\label{eq:psi_xi_def}
\Psi=\left(\begin{matrix}\lambda^\dag/\langle g\rangle\\ \psi^1_{\xi}\end{matrix}\right)\ , \quad \text{and}\quad \psi^1_{\xi}=\frac{ \langle g_{A\bar B}\bar k^{\bar B}\rangle \psi^A }{\sqrt{\langle g_{A\bar B}k^A\bar k^{\bar B}\rangle}}\ , \quad 
\psi_{\xi}^{1\dag}=\frac{ \langle g_{A\bar B} k^{A}\rangle \psi^{\bar B\dag} }{\sqrt{\langle g_{A\bar B}k^A\bar k^{\bar B}\rangle}}\ .
\end{split}
\end{equation}
We learned that when we evaluate the cross couplings in the newly defined vacuum, a single linear combination of fermions, called $\psi^1_\xi$, combines with the existing gaugino to form a massive Dirac spinor,
Examining eq. \eqref{eq:psi_xi_def}, we learn that the state $\psi^1_{\xi}$ is a linear combination of the $N$ fermions $\psi^A$. Note that we can use the same rotation matrix $U$, defined for the scalars, to express this linear combination.
\begin{equation}
\label{eq:both_rotations}
\xi^1=[U]^1_A\delta z^A\ , \quad \psi^1_{\xi}=[U]^1_A\psi^A\ .
\end{equation}
What this shows is that the fermion eigenstate $\psi^1_{\xi}$ corresponds to the scalar eigenstate $\xi^1$ They belong in the same chiral supermultiplet $(\xi^1,\psi^{1}_{\xi})$. This chiral supermultiplet becomes massive after the system is stabilized into a supersymmetric vacuum, but is absorbed into a massive vector multiplet. The imaginary component of $\xi^1$, $\phi$, becomes the real scalar component of this vector multiplet. We have also shown that $\eta$, the real component of $\xi^1$, becomes the longitudinal degree of freedom of the massive vector boson $A_\mu$. We are left to discuss the fermionic component of this vector multiplet, and show that it has the same mass as the field $\phi$, a direct consequence of unbroken supersymmetry.

We found a mass term for a Dirac fermion $\Psi$: $\sqrt{2 \langle g^2g_{A\bar B}k^A\bar k^{\bar B}\rangle}\bar \Psi \Psi$. To determine the mass of $\Psi$, we have to ensure its kinetic energy is canonically normalized. We have already seen that the fields $\xi^1$ and $\psi^1_{\xi}$ are defined with the same rotation matrix $U$. In fact, we form other $N-1$ chiral multiplets with the same rotation matrix. These eigenstates, however, pair into massless multiplets $(\xi_A,\psi_{A\xi})$, $i=2,\dots, N$. We write
\begin{equation}
\begin{split}
\label{eq:rest_states1}
&\xi^A=[U_B^A]\delta z^B\ , \quad A=2,\cdots,N\ ,\\
&\psi_\xi^A=[U]^A_B\psi^B\ ,\quad B=1,\cdots, N\ .
\end{split}
\end{equation}

 In terms of the new fermion eigenstates, the kinetic term of the chiral fermionic fields $\psi^i$ becomes
\begin{equation}
 -ig_{A\bar B}\psi^A  \slashed{\partial} \psi^{\dag \bar B}\rightarrow -i g_{A\bar B}[U^{-1}]^A_C[U^{*-1}]^{\bar B}_{\bar D}\psi^C_\xi
 \slashed{\partial}\psi_{\xi}^{\dag \bar D}=-i\delta_{C\bar D}\psi^C_\xi \slashed{\partial}\psi_{\xi}^{\dag \bar D}\ ,
\end{equation}
where we made use of the property of the matrix $U^{-1}$ of diagonalizing the metric, as shown in \eqref{eq:matrix_condition_x}. 
 We learn that the massive field $\psi^1_{\xi}$, as well as the massless fermions $\psi^A_{ \xi}$, for $A=2,\dots,N$ have canonically normalized kinetic terms, which was to be expected. We can therefore write the following Lagrangian for the fermions (neglecting interactions)
\begin{equation}
\begin{split}
\mathcal{L}\supset& -i\psi^1_\xi \slashed{\partial} \psi_\xi^{1\dag}  -i \psi_{A\xi} \slashed{\partial} \psi_\xi^{A\dag}-\tfrac{i}{g^2} \lambda \slashed{\partial} \lambda^\dag+\sqrt{2 \langle g^2 g_{A\bar B}k^A\bar k^{\bar B}\rangle}
\left(\psi^{1\dag}_\xi \frac{\lambda^{\dag}}{\langle g\rangle}+\psi^1_\xi\frac{\lambda}{\langle g\rangle} \right)\\
&=i\bar \Psi \slashed\partial \Psi  -i \psi_{A\xi} \slashed{\partial} \psi_\xi^{A\dag}+\sqrt{2 \langle g^2 g_{A\bar B}k^A\bar k^{\bar B}\rangle}\bar \Psi \Psi\ , \quad A=2,\dots, N
\end{split}
\end{equation}
This is the equation of motion for a Dirac fermion of the form $\Psi=\left( \begin{matrix} \lambda^\dag/\langle g\rangle\\\psi \end{matrix} \right)$, with mass $M_\Psi=\sqrt{2 \langle g^2 g_{A\bar B}k^A\bar k^{\bar B}\rangle}$, and $N-1$ massless Weyl spinors $\Psi^A_\xi$.

\end{itemize}
 
In conclusion, we found that after we fix the a linear combination of the VEVs of the scalar fields $z^A$ to obtain a supersymmetric vacuum, we obtain 
a massive vector multiplet, containing a real scalar, a massive vector boson and a Dirac fermion. That is,
\begin{equation}
\left( \phi, A_\mu, \Psi \right)\ .
\end{equation}
The real scalar and the fermions are linear combinations of fields which receive mass terms after we fix the VEVs of the scalars and expand around the vacuum:
\begin{equation}
\begin{split}
\phi=\frac{ i\langle k^Ag_{A\bar B}\rangle \tfrac{\delta \bar z^{\bar B}}{2}-i\langle k^{\bar B}g_{A\bar B}\rangle \tfrac{\delta z^A}{2}  }{\sqrt{ \langle g_{A\bar B}k^A\bar k^{\bar B}\rangle}}\\
\end{split}
\end{equation}

All components of the massive vector multiplet have the same mass,
\begin{equation}
\label{eq:masses_equal}
m_{A}=M_\Psi=m_\phi=\sqrt{ 2\langle  g^2 g_{A\bar B}k^A\bar k^{\bar B}\rangle}\ ,
\end{equation}
as expected, since we fixed the vacuum to be supersymmetric, with $V_D=0$.

Furthermore, we found that the effective theory contains another $N-1$ massless chiral multiplets  $(\xi^A, \psi^A)$, for $A=2,\cdots N$. These chiral components are given by the linear combinations given in \eqref{eq:rest_states1}.  The 
scalar components $\xi^A$, $A=2,\dots,N$ are linear combinations of the scalar perturbations around the vacuum, which, however, receive no mass terms. A D-term potential develops in the direction of the $\xi^A$ scalar only, while the rest of the states $\xi^A$, $A=2,\dots,N$, which are orthogonal to it, remain flat. 

Having computed all the mass eigenstates after the D-term stabilization, we are ready to express the complete Lagrangian of the low-energy theory. We find
\begin{equation}
\begin{split}
\mathcal{L}&\supset
-\partial_\mu \phi \partial^\mu \phi - m_\phi^2\phi^2+i\bar \Psi \slashed\partial \Psi+M_\Psi\bar \Psi \Psi\ 
 -\frac{1}{4g^2}F_{\mu\nu}F^{\mu \nu}-\frac{1}{2\langle g^2\rangle}m_A^2A_\mu A^{\mu} \\
 &-\sum_{A=2}^{N}\partial_\mu \xi_A\partial^\mu \bar \xi^A-\sum_{A=2}^{N}i\psi_{A\xi}\slashed{\partial}\psi^{A\dag}_\xi+\dots\ , \qquad A=2,\cdots, N\ .
 \end{split}
\end{equation}  
We have omitted the interaction terms. The masses $m_\phi$, $M_\Psi$ and $m_A$ are equal, as shown in eq. 
\eqref{eq:masses_equal}. The vector boson is defined in the "unitary" gauge.

\section{K\"ahler metrics}

The complete K\"ahler potential for the $S$, $T^i$, $Z$ and $C^L$ fields is given by
\begin{equation}
K=K_S+K_T+K_{\text{matter}}\ ,
\end{equation}
where
\begin{equation}
\begin{split}
\label{}
&K_S=-\kappa_4^{-2}\ln\left(S+\bar S-\frac{\pi}{2}\epsilon_S \frac{(Z+\bar Z)^2}{W_i(T^i+\bar T^i)}\right)\ , \\
&K_T=-\kappa_4^{-2}\ln\left(\frac{1}{48}d_{ijk}(T^i+\bar T^i)(T^j+\bar T^j)(T^k+\bar T^k)\right)\ ,\\
&K_{\text{matter}}=e^{ \kappa_4^2K_T/3}\mathcal{G}_{L\bar M}C^L\bar C^{\bar M}\ .
\end{split}
\end{equation}
The first derivatives of the K\"ahler potential with respect to $S$, $T^i$, $Z$ and $C^L$ are
\begingroup
\allowdisplaybreaks
\begin{align}
\frac{\partial K}{\partial S}=&\frac{\partial K_S}{\partial S}=-\frac{1}{\kappa_4^2\left(S+\bar S-\frac{\pi}{2}\epsilon_S \frac{(Z+\bar Z)^2}{W_i(T^i+\bar T^i)}\right)}=-\frac{1}{2\kappa^2_4V}\ ,\\
\frac{\partial K}{\partial T^i}
=&\frac{\partial K_T}{\partial T^i}    -\frac{\frac{\pi}{2}\epsilon_S \frac{W_i(Z+\bar Z)^2}{\left(W_j(T^j+\bar T^j)\right)^2}}{\kappa_4^2\left(S+\bar S-\frac{\pi}{2}\epsilon_S \frac{(Z+\bar Z)^2}{W_j(T^j+\bar T^j)}\right)}    +\frac{\kappa_4^2\partial K_T}{\partial T^i}\frac{e^{\kappa_4^2K_T/3}}{3}\mathcal{G}_{L\bar M}C^L\bar C^{\bar M}\\
=&-\frac{d_{ijk}a^ja^k}{4\kappa_4^2\hat RV^{2/3}}-\frac{1}{4\kappa_4^2V}\pi\epsilon_Sz^2W^i-\frac{d_{ijk}a^ja^k}{4\hat RV^{2/3}}\frac{e^{\kappa_4^2K_T/3}}{3}\mathcal{G}_{L\bar M}C^L\bar C^{\bar M}\ ,\\
\frac{\partial K}{\partial Z}=& \frac{\partial K_S}{\partial Z}= \frac{{\pi}\epsilon_S \frac{(Z+\bar Z)}{W_i(T^i+\bar T^i)}}{\kappa_4^2\left(S+\bar S-\frac{\pi}{2}\epsilon_S \frac{(Z+\bar Z)^2}{W_i(T^i+\bar T^i)}\right)}  =\frac{\pi\epsilon_Sz}{2\kappa^2_4V}\ ,\\
\frac{\partial K}{\partial C^L}=&\frac{\partial K_{\text{matter}}}{\partial C_L}=e^{\kappa_4^2K_T/3}\mathcal{G}_{L\bar M}\bar C^{\bar M}\ .
\end{align}
\endgroup

The second derivatives of the K\"ahler potential with respect to $S$, $T^i$, $Z$ and $C^L$ are
\begingroup
\allowdisplaybreaks
\begin{equation}
\begin{split}
\label{black1}
g_{S\bar S}=\frac{\partial^2 K}{\partial S\partial \bar S}&=\frac{1}{\kappa_4^2\left(S+\bar S-\frac{\pi}{2}\epsilon_S \frac{(Z+\bar Z)^2}{W_k(T^k+\bar T^k)}\right)^2}\\
&=\frac{1}{4\kappa^2_4V^2}\ ,\\
g_{T^i\bar S}=\frac{\partial^2K}{\partial T^i\partial \bar S}&=\frac{\frac{\pi}{2}\epsilon_SW_i \frac{(Z+\bar Z)^2}{\left(W_k(T^k+\bar T^k)\right)^2}}{\kappa_4^2\left(S+\bar S-\frac{\pi}{2}\epsilon_S \frac{(Z+\bar Z)^2}{W_k(T^k+\bar T^k)}\right)^2}  \\
&=\frac{1}{8\kappa_4^2V^2}\pi\epsilon_Sz^2W^i\ ,\\
g_{Z\bar S}=\frac{\partial^2K}{\partial Z\partial \bar S}&=-\frac{{\pi}\epsilon_S \frac{(Z+\bar Z)}{W_k(T^k+\bar T^k)}}{\kappa_4^2\left(S+\bar S-\frac{\pi}{2}\epsilon_S \frac{(Z+\bar Z)^2}{W_k(T^k+\bar T^k)}\right)^2} \\
& =-\frac{\pi\epsilon_Sz}{4\kappa^2_4V^2}\ ,\\
g_{T^i\bar Z}=\frac{\partial^2K}{\partial T^i\partial \bar Z}&=
- \frac{{\pi}\epsilon_SW_i \frac{(Z+\bar Z)}{\left(W_k(T^k+\bar T^k)\right)^2}}{\kappa_4^2\left(S+\bar S-\frac{\pi}{2}\epsilon_S \frac{(Z+\bar Z)^2}{W_k(T^k+\bar T^k)}\right)}
 -\frac{{\pi}\epsilon_S \frac{(Z+\bar Z)}{W_k(T^k+\bar T^k)}\frac{\pi}{2}\epsilon_S W_i\frac{(Z+\bar Z)^2}{\left(W_k(T^k+\bar T^k)\right)^2}
 }{\kappa_4^2\left(S+\bar S-\frac{\pi}{2}\epsilon_S \frac{(Z+\bar Z)^2}{W_k(T^k+\bar T^k)}\right)^2}\\
 &=-\frac{{\pi}\epsilon_SW_i{z}}{4\kappa_4^2V(W_kt^k)}-\frac{{\pi^2}\epsilon_S^2W_iz^3}{8\kappa_4^2V^2}\ ,\\ 
\end{split}
\end{equation}
 \begin{equation}
\begin{split}
g_{T^i\bar T^j}&=\frac{\partial^2K}{\partial T^i\partial \bar T^j}\\
&=\frac{\partial^2 K_T}{\partial T^i\partial \bar T^j}
  +\frac{{\pi}\epsilon_SW_iW_j \frac{(Z+\bar Z)^2}{\left(W_k(T^k+\bar T^k)\right)^3}}{\kappa_4^2\left(S+\bar S-\frac{\pi}{2}\epsilon_S \frac{(Z+\bar Z)^2}{W_k(T^k+\bar T^k)}\right)}  
    +\frac{\frac{\pi^2}{4}\epsilon_S^2 W_iW_j\frac{(Z+\bar Z)^4}{\left(W_k(T^k+\bar T^k)\right)^4}}{\kappa_4^2\left(S+\bar S-\frac{\pi}{2}\epsilon_S \frac{(Z+\bar Z)^2}{W_k(T^k+\bar T^k)}\right)^2} \\
    &+\frac{\kappa_4^2\partial^2 K_T}{\partial T^i\partial \bar T^j}\frac{e^{\kappa_4^2K_T/3}}{3}\mathcal{G}_{L\bar M}C^L\bar C^{\bar M}+\frac{\kappa_4^2\partial K_T}{\partial T^i}\frac{\kappa_4^2\partial K_T}{\partial \bar T^j}\frac{e^{\kappa_4^2K_T/3}}{9}\mathcal{G}_{L\bar M}C^L\bar C^{\bar M}\\
&=g^T_{ij}+\frac{\pi \epsilon_SW_iW_jz^2}{4\kappa_4^2V(W_kt^k)}
+\frac{\pi^2\epsilon_S^2W_iW_jz^4}{16\kappa_4^2V^2} +\left(\kappa_4^2g_{ij}^T+\frac{1}{3}\left(\frac{d_{ijk}a^ja^k}{4\hat RV^{2/3}}\right)^2\right)\frac{e^{\kappa_4^2K_T/3}}{3}\mathcal{G}_{L\bar M}C^L\bar C^{\bar M}\ ,\\
 g_{Z\bar Z}&=\frac{\partial K}{\partial Z\partial Z}=  \frac{{\pi}\epsilon_S \frac{1}{W_k(T^k+\bar T^k)}}{\kappa_4^2\left(S+\bar S-\frac{\pi}{2}\epsilon_S \frac{(Z+\bar Z)^2}{W_k(T^k+\bar T^k)}\right)} 
+\frac{{\pi}\epsilon_S \frac{(Z+\bar Z)}{W_k(T^k+\bar T^k)}{\pi}\epsilon_S \frac{(Z+\bar Z)}{W_k(T^k+\bar T^k)}}{\kappa_4^2\left(S+\bar S-\frac{\pi}{2}\epsilon_S \frac{(Z+\bar Z)^2}{W_k(T^k+\bar T^k)}\right)^2} 
,\\
&=\frac{{\pi}\epsilon_S}{4\kappa_4^2V}\frac{1}{W_kt^k}+\frac{\pi^2\epsilon_S^2z^2}{4\kappa_4^2V^2}\\
 \nonumber g_{T^i\bar C^{L}}&=\frac{\partial^2K}{\partial T^i\partial \bar C^{\bar L}}=\frac{1}{3}\frac{\kappa_4^2\partial K_T}{\partial T^i}e^{\kappa_4^2K_T/3}\mathcal{G}_{L\bar M}\bar C^{\bar M}\\
&=\frac{d_{ijk}a^ja^k}{12\hat RV^{2/3}}e^{\kappa_4^2K_T/3}\mathcal{G}_{L\bar M}\bar C^{\bar M}\ ,\\
  g_{C^L\bar C^{\bar M}}&=-\frac{\partial^2K}{\partial C^L\partial \bar C^{\bar M}}=e^{\kappa_4^2K_T/3}\mathcal{G}_{L\bar M}\ .\\
\end{split}
\end{equation}
\endgroup
In the above, we have defined
\begin{equation}
g_{i\bar j}^T=\frac{\partial^2 K_T}{\partial T^i\partial {\bar T^{\bar j}}}=-\frac{d_{ijk}t^k}{\kappa_4^2(\tfrac{2}{3})d_{lmn}t^lt^mt^n}+\frac{d_{ikl}t^kt^ld_{jmn}t^mt^n}{\kappa_4^2\left(\tfrac{2}{3}\right)^2(d_{lmn}t^lt^mt^n)^2}\ .
\end{equation}
Of course, we also have
\begin{equation}
g_{S\bar T^i}=g_{T^i\bar S}\ ,\quad g_{S\bar Z}=g_{Z\bar S}\ ,\quad g_{Z\bar T^i}=g_{T^i\bar Z}\ ,\quad \text{and}\quad g_{C_L\bar T^i}=g^*_{T^i\bar C^{\bar L}}\ .
\end{equation}

\end{document}